\documentclass{article}%
\usepackage{amsmath}
\usepackage{amsfonts}
\usepackage{amssymb}
\usepackage{graphicx}%
\newtheorem{theorem}{Theorem}

\newtheorem{proposition}[theorem]{Proposition}
\newtheorem{remark}[theorem]{Remark}

\newcommand{\bpartial}{\mathop{\partial\kern -4pt\raisebox{.8pt}{$|$}}}
\newcommand{\bra}{\mathopen{[\kern-1.6pt[}}
\newcommand{\ket}{\mathclose{]\kern-1.5pt]}}
\newcommand{\bbra}{\mathopen{[\kern-2.2pt[\kern-2.3pt[}}
\newcommand{\bket}{\mathclose{]\kern-2.1pt]\kern-2.3pt]}}

\newcommand{\slx}{\mbox{\bfseries\slshape x}}
\newcommand{\sly}{\mbox{\bfseries\slshape y}}

\makeindex
\begin{document}

\title{The Effective Lorentzian and Teleparallel Spacetimes Generated by a Free
Electromagnetic Field}
\author{E. Notte-Cuello$^{(1)}$, R. da Rocha$^{(2)}$ and W. A. Rodrigues Jr.$^{(3)}$\\$^{(1)}${\small Departamento de} {\small Matem\'{a}ticas,}\\{\small Universidad de La Serena}\\{\small Av. Cisternas 1200, La Serena-Chile}\\$^{(2)}${\footnotesize Instituto de F\'{\i}sica Te\'{o}rica, UNESP, Rua
Pamplona 145, 01405-900, S\~{a}o Paulo, SP, Brazil.}\\{\footnotesize and}\\{\footnotesize DRCC - Institute of Physics Gleb Wataghin, UNICAMP CP 6165}\\{\footnotesize 13083-970 Campinas, SP, Brazil}\\$^{(3)}\hspace{-0.1cm}${\footnotesize Institute of Mathematics, Statistics and
Scientific Computation}\\{\footnotesize IMECC-UNICAMP CP 6065}\\{\footnotesize 13083-859 Campinas, SP, Brazil}\\{\footnotesize \texttt{walrod@ime.unicamp.br; {\small enotte@userena.cl;
}roldao@ifi.unicamp.br}}}
\maketitle

\begin{abstract}
In this paper we show that a free electromagnetic field living in Minkowski
spacetime generates an effective Weitzenb\"{o}ck or an effective Lorentzian
spacetime whose properties are determined in details. These results are
possible because we found using the Clifford bundle formalism the noticeable
result that the energy-momentum densities of a free electromagnetic field are
sources of the Hodge duals of exact $2$-form fields which satisfy Maxwell like equations.

\end{abstract}

\section{Introduction}

In this paper we determine an effective Weitzenb\"{o}ck \ [$(M,\mathbf{g}%
,\nabla,\uparrow,\tau_{\mathtt{g}})$] spacetime and an effective Lorentzian
[$(M,\mathbf{g},\overset{\mathbf{g}}{D},\uparrow,\tau_{\mathtt{g}})$]
spacetime determined by a free electromagnetic field $F\in\sec%
{\displaystyle\bigwedge\nolimits^{2}}
T^{\ast}M$ configuration living on Minkowski spacetime [$(M,\mathbf{\eta
},D,\uparrow,\tau_{\mathtt{\eta}})$] and satisfying Maxwell equation $%
\bpartial
$ $F=0$. The determination of these effective spacetimes become easily
feasible because using the Clifford bundle formalism we found that the
energy-momentum densities $\mathcal{T}_{\mathbf{a}}\in\sec\bigwedge
\nolimits^{1}T^{\ast}M$ of the Maxwell field are sources of exact $2$-forms
fields $\mathcal{W}^{\mathbf{a}}\in\sec%
{\displaystyle\bigwedge\nolimits^{2}}
T^{\ast}M$ (which we explicitly identify) and which satisfy Maxwell like
equations $d\mathcal{W}^{\mathbf{a}}=0,$ $\underset{\mathtt{\eta}}{\delta
}\mathcal{W}^{\mathbf{a}}=-\mathcal{T}^{\mathbf{a}}$ in Minkowski spacetime.
Our results depends crucially on the Riesz formula $\mathcal{T}_{\mathbf{a}%
}=\frac{1}{2}F\vartheta_{\mathbf{a}}\tilde{F}$, which, of course has meaning
only in the Clifford bundle formalism. \ We identify $\mathbf{g}%
=\mathtt{h}^{\ast}\mathbf{\eta}$ \ as a metric field on $M$ generated by a
diffeomorphism $\mathtt{h}:M\rightarrow M$ associated with a distortion in the
Minkowski vacuum caused by $A\in\sec%
{\displaystyle\bigwedge\nolimits^{1}}
T^{\ast}M$ \ ($F=dA$). In the case of the effective Weitzenb\"{o}ck
(teleparallel) spacetime we give the explicit form for the torsion $2$-forms
$\Theta^{\mathbf{a}}$ and show that their sources are $T^{\mathbf{a}%
}=\mathtt{h}^{\ast}\mathcal{T}^{\mathbf{a}}$, more precisely, $d\underset
{\mathtt{g}}{\star}\Theta^{\mathbf{a}}=-\underset{\mathtt{g}}{\star
}T^{\mathbf{a}}$. In the case of the effective Lorentzian spacetime we give
the explicit forms of the Ricci $1$-form fields $\mathcal{R}^{\mathbf{a}}%
\in\sec\bigwedge\nolimits^{1}T^{\ast}M$ which can also be expressed in terms
of the Hodge Laplacian and covariant D'Alembertian of the $\mathbf{g}%
$-orthonormal cotetrads $\{\theta^{\mathbf{a}}\}$. The explicit formulas for
the $\mathcal{R}^{\mathbf{a}}$ for our problem gives in one sense an unified
theory of the gravitational and electromagnetic field in the sense of the
Rainich-Misner-Wheeler theory. \ The details of the present paper are as
follows. In Section 2 we present Maxwell equation in the Clifford bundle
$\mathcal{C}\!\ell(M,\mathtt{\eta})$, prove Riesz formula $\mathcal{T}%
_{\mathbf{a}}=\frac{1}{2}F\vartheta_{\mathbf{a}}\tilde{F}$ and obtain the
Maxwell like equations $%
\bpartial
$ $\mathcal{W}^{\mathbf{a}}=\mathcal{T}^{\mathbf{a}}$. In Section 3 we derive
the effective Weitzenb\"{o}ck (teleparallel) spacetime generated by $F$ and in
Section 4 we derive the effective Lorentzian spacetime generated by $F$. \ In
Section 5 we present our conclusions. The paper has several
Appendices\footnote{If they are not enough for completely intelligibility of
the paper, please consult \cite{rodcap}.}. Appendix A recalls the definition
of Clifford bundles. In Appendix A.1 the Clifford product is introduced and we
present several Clifford algebra identities necessary to follows the
calculations. Next in Appendix A.2 we introduce the definition of the Hodge
star operator associated with the metrics $\mathbf{\eta}$ and $\mathbf{g=}%
\mathtt{h}^{\ast}\mathbf{\eta}$ and show to calculate each one it using
Clifford algebra methods and also how they are related. The important notion
of the Dirac operator associated with a given Levi-Civita connection of a
metric field is introduced in Appendix A.5.

\section{Maxwell Equation}

We start this section by recalling that Eq.(\ref{14bis}) of Appendix permits
us to write the Maxwell equations
\begin{equation}
dF=0,\text{ }\delta F=-J \label{14}%
\end{equation}
for $F\in\sec%
{\displaystyle\bigwedge\nolimits^{2}}
T^{\ast}M\hookrightarrow\sec\mathcal{C}\!\ell(M,\mathtt{\eta})$\ living in
Minkowski spacetime as a \textit{single} equation (Maxwell equation),%
\begin{equation}%
\bpartial
\text{ }F=J. \label{Maxwell}%
\end{equation}

Next we investigate a noticeable formula which can be written only in the
Clifford bundle formalism and which is essential for all our developments.

\subsection{The Noticeable Riesz Formula $\mathcal{T}_{\mathbf{a}}=\frac{1}%
{2}F\vartheta_{\mathbf{a}}\tilde{F}$}

We now prove that the energy-momentum densities $\underset{\mathbf{\eta}%
}{\star}\mathcal{T}_{\mathbf{a}}$ of the Maxwell field can be written in the
Clifford bundle formalism as\footnote{The formula $\mathcal{T}_{\mathbf{a}%
}=\frac{1}{2}F\vartheta_{\mathbf{a}}\tilde{F}$ has been first obtained (but,
not using the algebraic derivatives of the Lagrangian density) by M. Riesz in
1947 \cite{riesz} and it has been rediscovered by Hestenes in 1996
\cite{hestenes} (which also does not use the algebraic derivatives of the
Lagrangian density). Algebraic derivatives of \textit{homogenous} form fields
has been described, e.g., in Thirring's book \cite{thirring}.}:
\begin{equation}
\underset{\mathbf{\eta}}{\star}\mathcal{T}_{\mathbf{a}}=\frac{1}{2}%
\underset{\mathbf{\eta}}{\star}(F\vartheta_{\mathbf{a}}\tilde{F})\in
\sec\bigwedge\nolimits^{3}T^{\ast}M\hookrightarrow\sec\mathcal{C\ell
(}M,\mathtt{\eta)}. \label{em1}%
\end{equation}

To derive Eq.(\ref{em1}) we start from the Maxwell Lagrangian
\begin{equation}
\mathcal{L}_{m}=\frac{1}{2}F\wedge\underset{\mathbf{\eta}}{\star}F,
\label{em2}%
\end{equation}
where $F=\frac{1}{2}F_{\mathbf{ab}}\vartheta^{\mathbf{a}}\wedge\vartheta
^{\mathbf{b}}:=\frac{1}{2}F_{\mathbf{ab}}\vartheta^{\mathbf{ab}}\in\sec%
{\displaystyle\bigwedge\nolimits^{2}}
TM\hookrightarrow\sec\mathcal{C\ell(}M,\mathtt{\eta)}$ is the electromagnetic
field. Now, denoting by $%
\mbox{\boldmath{$\delta$}}%
$ the variational symbol\footnote{Please, do not confuse the variational
symbol $%
\mbox{\boldmath{$\delta$}}%
$ with the symbol $\delta$ of the Hodge coderiviative.} we can easily verify
that
\[%
\mbox{\boldmath{$\delta$}}%
\underset{\mathbf{\eta}}{\star}\vartheta^{\mathbf{ab}}=%
\mbox{\boldmath{$\delta$}}%
\theta^{\mathbf{c}}\wedge\lbrack\vartheta_{\mathbf{c}}\underset{\mathtt{\eta}%
}{\lrcorner}\underset{\mathbf{\eta}}{\star}\vartheta^{\mathbf{ab}}].
\]
Moreover, in general $%
\mbox{\boldmath{$\delta$}}%
$ and $\underset{\mathbf{\eta}}{\star}$ do not commute. Indeed,\ for any
\ $\mathcal{A}_{p}\in\sec%
{\displaystyle\bigwedge\nolimits^{p}}
T^{\ast}M\hookrightarrow\sec\mathcal{C\ell(}M,\mathtt{\eta)}$ we have
\begin{align}
\lbrack%
\mbox{\boldmath{$\delta$}}%
,\underset{\mathbf{\eta}}{\star}]\mathcal{A}_{p}  &  =%
\mbox{\boldmath{$\delta$}}%
\underset{\mathbf{\eta}}{\star}\mathcal{A}_{p}-\underset{\mathbf{\eta}}{\star}%
\mbox{\boldmath{$\delta$}}%
\mathcal{A}_{p}\label{7.ex00}\\
&  =%
\mbox{\boldmath{$\delta$}}%
\vartheta^{\mathbf{a}}\wedge\left(  \vartheta_{\mathbf{a}}\underset
{\mathtt{\eta}}{\lrcorner}\underset{\mathbf{\eta}}{\star}\mathcal{A}%
_{p}\right)  -\underset{\mathbf{\eta}}{\star}\left[
\mbox{\boldmath{$\delta$}}%
\vartheta^{\mathbf{a}}\wedge\left(  \vartheta_{\mathbf{a}}\underset
{\mathtt{\eta}}{\lrcorner}\mathcal{A}_{p}\right)  \right]  .\nonumber
\end{align}
Multiplying both members of Eq.(\ref{7.ex00}) with $\mathcal{A}_{p}=F$ on the
right by $F\wedge$ we get%
\[
F\wedge%
\mbox{\boldmath{$\delta$}}%
\underset{\mathbf{\eta}}{\star}F=F\wedge\underset{\mathbf{\eta}}{\star}%
\mbox{\boldmath{$\delta$}}%
F+F\wedge\{%
\mbox{\boldmath{$\delta$}}%
\vartheta^{\mathbf{a}}\wedge(\vartheta_{\mathbf{a}}\underset{\mathtt{\eta}%
}{\lrcorner}\underset{\mathbf{\eta}}{\star}F)-\underset{\mathbf{\eta}}{\star}[%
\mbox{\boldmath{$\delta$}}%
\vartheta^{\mathbf{a}}\wedge(\vartheta_{\mathbf{a}}\underset{\mathtt{\eta}%
}{\lrcorner}F)]\}.
\]

Next we sum $%
\mbox{\boldmath{$\delta$}}%
F\wedge\underset{\mathbf{\eta}}{\star}F$ to both members of the above equation obtaining%

\[%
\mbox{\boldmath{$\delta$}}%
\left(  F\wedge\underset{\mathbf{\eta}}{\star}F\right)  =2%
\mbox{\boldmath{$\delta$}}%
F\wedge\underset{\mathbf{\eta}}{\star}F+%
\mbox{\boldmath{$\delta$}}%
\vartheta^{\mathbf{a}}\wedge\lbrack F\wedge(\vartheta_{\mathbf{a}}%
\underset{\mathtt{\eta}}{\lrcorner}\underset{\mathbf{\eta}}{\star
}F)-(\vartheta_{\mathbf{a}}\underset{\mathtt{\eta}}{\lrcorner}F)\wedge
\underset{\mathbf{\eta}}{\star}F].
\]

Then, it follows (see, \cite{rodcap,quiroro} for some details) that
if\footnote{$\pounds _{\xi}$ denotes the Lie derivative in the direction of
the vector field $\xi$.} $%
\mbox{\boldmath{$\delta$}}%
\vartheta^{\mathbf{a}}=-\pounds _{\xi}\vartheta^{\mathbf{a}}$, for some
diffeomorphism generated by the vector field $\xi$ that
\[
\underset{\mathbf{\eta}}{\star}\mathcal{T}_{\mathbf{a}}=\frac{\partial
\mathcal{L}_{m}}{\partial\vartheta^{\mathbf{a}}}=\frac{1}{2}\left[
F\wedge(\vartheta_{\mathbf{a}}\underset{\mathtt{\eta}}{\lrcorner}%
\underset{\mathtt{\eta}}{\star}F)-(\vartheta_{\mathbf{a}}\underset
{\mathtt{\eta}}{\lrcorner}F)\wedge\underset{\mathtt{\eta}}{\star}F\right]  .
\]
Now,
\[
(\vartheta_{\mathbf{a}}\underset{\mathtt{\eta}}{\lrcorner}F)\wedge
\underset{\mathtt{\eta}}{\star}F=-\underset{\mathtt{\eta}}{\star}%
[(\vartheta_{\mathbf{a}}\underset{\mathtt{\eta}}{\lrcorner}F)\underset
{\mathtt{\eta}}{\lrcorner}F]=-[(\vartheta_{\mathbf{a}}\underset{\mathtt{\eta}%
}{\lrcorner}F)\underset{\mathtt{\eta}}{\lrcorner}F]\tau_{\mathtt{\mathbf{\eta
}}}%
\]
and using also the identity \cite{rodcap}
\[
(\vartheta_{\mathbf{a}}\underset{\mathtt{\eta}}{\lrcorner}F)\wedge
\underset{\mathtt{\eta}}{\star}F=\vartheta_{\mathbf{a}}(F\underset
{\mathtt{\eta}}{\cdot}F)\tau_{\mathtt{\eta}}-F\wedge(\vartheta_{\mathbf{a}%
}\underset{\mathtt{\eta}}{\lrcorner}\underset{\mathtt{\eta}}{\star}F).
\]
we get%

\begin{align*}
\frac{1}{2}\left[  F\wedge(\vartheta_{\mathbf{a}}\underset{\mathtt{\eta}%
}{\lrcorner}\underset{\mathtt{\eta}}{\star}F)-(\vartheta_{\mathbf{a}}\lrcorner
F)\wedge\underset{\mathtt{\eta}}{\star}F\right]   &  =\frac{1}{2}\left\{
\vartheta_{\mathbf{a}}(F\underset{\mathtt{\eta}}{\cdot}F)\tau
_{\mathtt{\mathbf{\eta}}}-(\vartheta_{\mathbf{a}}\underset{\mathtt{\eta}%
}{\lrcorner}F)\wedge\underset{\mathtt{\eta}}{\star}F-(\vartheta_{\mathbf{a}%
}\underset{\mathtt{\eta}}{\lrcorner}F)\wedge\underset{\mathtt{\eta}}{\star
}F\right\} \\
&  =\frac{1}{2}\left\{  \vartheta_{\mathbf{a}}(F\underset{\mathtt{\eta}}%
{\cdot}F)\tau_{\mathtt{\mathbf{\eta}}}-2(\vartheta_{\mathbf{a}}\underset
{\mathtt{\eta}}{\lrcorner}F)\wedge\underset{\mathtt{\eta}}{\star}F\right\} \\
&  =\frac{1}{2}\left\{  \vartheta_{\mathbf{a}}(F\underset{\mathtt{\eta}}%
{\cdot}F)\tau_{\mathtt{\mathbf{\eta}}}+2[(\vartheta_{\mathbf{a}}%
\underset{\mathtt{\eta}}{\lrcorner}F)\underset{\mathtt{\eta}}{\lrcorner}%
F]\tau_{\mathtt{\mathbf{\eta}}}\right\} \\
&  =\underset{\mathtt{\eta}}{\star}\left(  \frac{1}{2}\vartheta_{\mathbf{a}%
}(F\underset{\mathtt{\eta}}{\cdot}F)+(\vartheta_{\mathbf{a}}\underset
{\mathtt{\eta}}{\lrcorner}F)\underset{\mathtt{\eta}}{\lrcorner}F\right)
=\frac{1}{2}\underset{\mathtt{\eta}}{\star}(F\vartheta_{\mathbf{a}}\tilde{F}),
\end{align*}
where in writing the last line we used the identity
\begin{equation}
\frac{1}{2}Fn\tilde{F}=(n\underset{\mathtt{\eta}}{\lrcorner}F)\underset
{\mathtt{\eta}}{\lrcorner}F+\frac{1}{2}n(F\underset{\mathtt{\eta}}{\cdot}F),
\label{6.66a}%
\end{equation}
whose proof is as follows:
\begin{align*}
(n\underset{\mathtt{\eta}}{\lrcorner}F)\underset{\mathtt{\eta}}{\lrcorner
}F+\frac{1}{2}n(F\underset{\mathtt{\eta}}{\cdot}F)  &  =\frac{1}{2}\left[
(n\underset{\mathtt{\eta}}{\lrcorner}F)F-F(n\underset{\mathtt{\eta}}%
{\lrcorner}F)\right]  +\frac{1}{2}n(F\underset{\mathtt{\eta}}{\cdot}F)\\
&  =\frac{1}{4}\left[  nFF-FnF-FnF+FFn\right]  +\frac{1}{2}n(F\underset
{\mathtt{\eta}}{\cdot}F)\\
&  =-\frac{1}{2}FnF+\frac{1}{4}\left[  -2n(F\underset{\mathtt{\eta}}{\cdot
}F)+n(F\wedge F)+(F\wedge F)n\right] \\
&  +\frac{1}{2}n(F\underset{\mathtt{\eta}}{\cdot}F)\\
&  =-\frac{1}{2}FnF-\frac{1}{2}n(F\underset{\mathtt{\eta}}{\cdot}F)+\frac
{1}{2}n\wedge(F\wedge F)+\frac{1}{2}n(F\underset{\mathtt{\eta}}{\cdot}F)\\
&  =-\frac{1}{2}FnF=\frac{1}{2}Fn\tilde{F}.
\end{align*}
valid for any $n\in\sec%
{\displaystyle\bigwedge\nolimits^{1}}
T^{\ast}M\hookrightarrow\sec\mathcal{C\ell}(M,\mathtt{\eta})$ \ and \ $F$
$\in\sec%
{\displaystyle\bigwedge\nolimits^{2}}
T^{\ast}M\hookrightarrow\sec\mathcal{C\ell}(M,\mathtt{\eta})$.

For completeness and presentation of \ some more tricks of the trade we detail
the proof that $\mathcal{T}_{\mathbf{a}}\cdot\vartheta_{\mathbf{b}%
}=\mathcal{T}_{\mathbf{b}}\cdot\vartheta_{\mathbf{a}}.$
\begin{align*}
\mathcal{T}_{\mathbf{a}}\underset{\mathtt{\eta}}{\cdot}\vartheta_{\mathbf{b}}
&  =-\frac{1}{2}\langle F\vartheta_{\mathbf{a}}F\vartheta_{\mathbf{b}}%
\rangle_{0}=-\langle(F\underset{\mathtt{\eta}}{\llcorner}\vartheta
_{\mathbf{a}})F\vartheta_{\mathbf{b}}\rangle_{0}-\frac{1}{2}\langle
(\vartheta_{\mathbf{a}}\underset{\mathtt{\eta}}{\lrcorner}F\text{ }%
+\vartheta_{\mathbf{a}}\wedge F)\text{ }F\vartheta_{\mathbf{b}}\rangle_{0}\\
&  =-\langle(F\underset{\mathtt{\eta}}{\llcorner}\vartheta_{\mathbf{a}%
})F\theta_{\mathbf{b}}\rangle_{0}-\frac{1}{2}\langle(\vartheta_{\mathbf{a}%
}FF\vartheta_{\mathbf{b}}\rangle_{0}\\
&  =-\langle(F\underset{\mathtt{\eta}}{\llcorner}\vartheta_{\mathbf{a}%
})(F\underset{\mathtt{\eta}}{\llcorner}\theta_{\mathbf{b}})+(F\underset
{\mathtt{\eta}}{\llcorner}\vartheta_{\mathbf{a}})(F\wedge\vartheta
_{\mathbf{b}})\rangle_{0}\\
&  +\frac{1}{2}\langle\vartheta_{\mathbf{a}}(F\underset{\mathtt{\eta}}{\cdot
}F)\vartheta_{\mathbf{b}}\rangle_{0}-\frac{1}{2}\langle\vartheta_{\mathbf{a}%
}(F\wedge F)\vartheta_{\mathbf{b}}\rangle_{0}\\
&  =-\langle(F\underset{\mathtt{\eta}}{\llcorner}\vartheta_{\mathbf{a}%
})(F\underset{\mathtt{\eta}}{\llcorner}\vartheta_{\mathbf{b}})\rangle
_{0}+\frac{1}{2}\langle(F\underset{\mathtt{\eta}}{\cdot}F)(\vartheta
_{\mathbf{a}}\underset{\mathtt{\eta}}{\cdot}\vartheta_{\mathbf{b}})\rangle
_{0}\\
&  =-(F\underset{\mathtt{\eta}}{\llcorner}\vartheta_{\mathbf{b}}%
)\cdot(F\underset{\mathtt{\eta}}{\llcorner}\vartheta_{\mathbf{a}})+\frac{1}%
{2}(F\underset{\mathtt{\eta}}{\cdot}F)(\vartheta_{\mathbf{b}}\underset
{\mathtt{\eta}}{\cdot}\vartheta_{\mathbf{a}})=\mathcal{T}_{\mathbf{b}%
}\underset{\mathtt{\eta}}{\cdot}\vartheta_{\mathbf{a}}.
\end{align*}

Note moreover that%
\begin{equation}
\mathcal{T}_{\mathbf{ab}}=\mathcal{T}_{\mathbf{a}}\underset{\mathtt{\eta}%
}{\cdot}\vartheta_{\mathbf{b}}=-\eta^{\mathbf{cl}}F_{\mathbf{ac}%
}F_{\mathbf{bl}}+\frac{1}{4}F_{\mathbf{cd}}F^{\mathbf{cd}}\eta_{\mathbf{ab}},
\end{equation}
a well known result.

Of course, for the \textit{free} electromagnetic field we have that
$d\underset{\mathbf{\eta}}{\star}\mathcal{T}^{\mathbf{a}}=0$, which is
equivalent to $\delta\mathcal{T}^{\mathbf{a}}=-%
\bpartial
{\underset{\mathtt{\eta}}{\lrcorner}\mathcal{T}^{\mathbf{a}}=0}$. Indeed,
observe that
\begin{align}%
\bpartial
{\underset{\mathtt{\eta}}{\lrcorner}\mathcal{T}^{\mathbf{a}}}{=}%
\bpartial
{\lrcorner}  &  \frac{1}{2}(F\vartheta^{\mathbf{a}}\tilde{F})\nonumber\\
&  =\frac{1}{2}\langle%
\bpartial
(F\vartheta^{\mathbf{a}}\tilde{F})\rangle_{0}\nonumber\\
&  =\frac{1}{2}\langle(%
\bpartial
F)\vartheta^{\mathbf{a}}\tilde{F}+\vartheta^{\mathbf{b}}\left(  F\vartheta
^{\mathbf{a}}D_{e_{\mathbf{b}}}\tilde{F}\right)  \rangle_{0}\nonumber\\
&  =\frac{1}{2}\langle\vartheta^{\mathbf{b}}\left(  F\vartheta^{\mathbf{a}%
}D_{e_{\mathbf{b}}}\tilde{F}\right)  \rangle_{0},
\end{align}
where we used that $%
\bpartial
$ $F=0$. Now,%

\begin{equation}%
\begin{array}
[c]{ll}%
\vartheta^{\mathbf{b}}\left(  F\vartheta^{\mathbf{a}}D_{e_{\mathbf{b}}}%
\tilde{F}\right)  & =\vartheta^{\mathbf{b}}\left\langle F\vartheta
^{\mathbf{a}}D_{e_{\mathbf{b}}}\tilde{F}\right\rangle _{1}+\vartheta
^{\mathbf{b}}\left\langle F\vartheta^{\mathbf{a}}D_{e_{\mathbf{b}}}\tilde
{F}\right\rangle _{3}\\
& =\vartheta^{\mathbf{b}}\lrcorner\left\langle F\vartheta^{\mathbf{a}%
}D_{e_{\mathbf{b}}}\tilde{F}\right\rangle _{1}+\vartheta^{\mathbf{b}}%
\wedge\left\langle F\vartheta^{\mathbf{a}}D_{e_{\mathbf{b}}}\tilde
{F}\right\rangle _{1}\\
& +\vartheta^{\mathbf{b}}\lrcorner\left\langle F\vartheta^{\mathbf{a}%
}D_{e_{\mathbf{b}}}\tilde{F}\right\rangle _{3}+\vartheta^{\mathbf{b}}%
\wedge\left\langle F\vartheta^{\mathbf{a}}D_{e_{\mathbf{b}}}\tilde
{F}\right\rangle _{3}.
\end{array}
\end{equation}
Then%
\begin{align*}
\langle\vartheta^{\mathbf{b}}\left(  F\vartheta^{\mathbf{a}}D_{e_{\mathbf{b}}%
}\tilde{F}\right)  \rangle_{0}  &  =\vartheta^{\mathbf{b}}\lrcorner
\left\langle F\vartheta^{\mathbf{a}}D_{e_{\mathbf{b}}}\tilde{F}\right\rangle
_{1}=\left\langle F\vartheta^{\mathbf{a}}D_{e_{\mathbf{b}}}\tilde
{F}\right\rangle _{1}\llcorner\vartheta^{\mathbf{b}}\\
&  =\left\langle F\vartheta^{\mathbf{a}}D_{e_{\mathbf{b}}}\tilde{F}%
\vartheta^{\mathbf{b}}\right\rangle _{0}\\
&  =\langle F\vartheta^{\mathbf{a}}\widetilde{(%
\bpartial
F)}\rangle_{0}=0,
\end{align*}
where we used the symbol $\widetilde{(%
\bpartial
F)}:=D_{e_{\mathbf{b}}}\tilde{F}\vartheta^{\mathbf{b}}$ and the fact that
\ $\widetilde{(%
\bpartial
F)}=0$.

\subsection{ Enter New Maxwell Like Equations $d\mathcal{W}^{\mathbf{a}}=0$,
$\underset{\eta}{\delta}\mathcal{W}^{\mathbf{a}}=-\mathcal{T}^{\mathbf{a}}$}

Let $\underset{\mathbf{\eta}}{\star}\mathcal{T}^{\mathbf{a}}=\frac{1}%
{2}\underset{\mathbf{\eta}}{\star}(F\vartheta^{\mathbf{a}}\tilde{F})\in\sec%
{\textstyle\bigwedge\nolimits^{3}}
T^{\ast}M\hookrightarrow\sec\mathcal{C\ell}(M,\mathtt{\eta})$ be the
energy-momentum densities of a free electromagnetic field configuration
$F\in\sec%
{\textstyle\bigwedge\nolimits^{2}}
T^{\ast}M\hookrightarrow\sec\mathcal{C\ell}(M,\mathtt{\eta})$ $(%
\bpartial
F=0)$. As we already know, we have
\begin{equation}
-\underset{\mathtt{\eta}}{\delta}\mathcal{T}^{\mathbf{a}}=%
\bpartial
\underset{\mathtt{\eta}}{{\lrcorner}}\mathcal{T}^{\mathbf{a}}=0. \label{J1}%
\end{equation}
Eq.(\ref{J1}) which is equivalent to $d\underset{\mathbf{\eta}}{\star
}\mathcal{T}^{\mathbf{a}}=0$ and since we are in Minkowski spacetime there
must exist $\mathcal{W}^{\mathbf{a}}\in\sec%
{\textstyle\bigwedge\nolimits^{2}}
T^{\ast}M\hookrightarrow\sec\mathcal{C\ell}(M,\mathtt{\eta})$ such that
\begin{equation}
-\mathcal{T}^{\mathbf{a}}=\text{ }\underset{\mathtt{\eta}}{\delta}%
\mathcal{W}^{\mathbf{a}} \label{the equation}%
\end{equation}

We claim that

\begin{proposition}%
\begin{align}
\mathcal{W}^{\mathbf{a}}  &  =d\Gamma^{\mathbf{a}},\label{J2}\\
\Gamma^{\mathbf{a}}  &  =-\frac{1}{4}(A\vartheta^{\mathbf{a}}A+X^{\mathbf{a}})
\end{align}
where $A$ is the electromagnetic potential $(dA=F)$, and $\ X^{\mathbf{a}}%
\in\sec%
{\textstyle\bigwedge\nolimits^{1}}
T^{\ast}M\hookrightarrow\sec\mathcal{C\ell}(M,\mathtt{\eta})$ and
\begin{equation}
dX^{\mathbf{a}}=-2\vartheta^{\mathbf{b}}\underset{\mathtt{\eta}}{\lrcorner
}\left\langle A\vartheta^{\mathbf{a}}D_{e_{\mathbf{b}}}A\right\rangle
_{3}+2B^{\mathbf{a}}, \label{J22}%
\end{equation}
with%

\begin{equation}%
\bpartial
\left\langle A\vartheta^{\mathbf{a}}F\right\rangle _{0}=%
\bpartial
\underset{\mathtt{\eta}}{\lrcorner}B^{\mathbf{a}}. \label{J23}%
\end{equation}

\end{proposition}

\noindent\textbf{Proof}: To prove the proposition we first note that
\[%
\begin{array}
[c]{ll}%
A\vartheta^{\mathbf{a}}A & =\left(  \vartheta^{\mathbf{a}}\underset
{\mathtt{\eta}}{\cdot}A\right)  A+A\left(  \vartheta^{\mathbf{a}}\wedge
A\right) \\
& =\left(  \vartheta^{\mathbf{a}}\underset{\mathtt{\eta}}{\cdot}A\right)
A+A\underset{\mathtt{\eta}}{{\lrcorner}}\left(  \vartheta^{\mathbf{a}}\wedge
A\right)  +A\wedge\left(  \vartheta^{\mathbf{a}}\wedge A\right)  ,
\end{array}
\]
and since $A\wedge\vartheta^{\mathbf{a}}\wedge A=0$ we have%
\begin{equation}
A\vartheta^{\mathbf{a}}A=A\vartheta^{\mathbf{a}}\widetilde{A}=\left\langle
A\vartheta^{\mathbf{a}}\widetilde{A}\right\rangle _{1}. \label{J3}%
\end{equation}
Then
\begin{align}
d\left(  A\vartheta^{\mathbf{a}}\widetilde{A}\right)   &  =d\left\langle
A\vartheta^{\mathbf{a}}\widetilde{A}\right\rangle _{1}=%
\bpartial
{\wedge}\left\langle A\vartheta^{\mathbf{a}}A\right\rangle _{1}=\vartheta
^{\mathbf{b}}\wedge\left\langle D_{e_{\mathbf{b}}}\left(  A\vartheta
^{\mathbf{a}}A\right)  \right\rangle _{1}\label{J4}\\
&  =\left\langle \left(  \vartheta^{\mathbf{b}}D_{e_{\mathbf{b}}}A\right)
\vartheta^{\mathbf{a}}\widetilde{A}+\vartheta^{\mathbf{b}}A\vartheta
^{\mathbf{a}}D_{e_{\mathbf{b}}}\widetilde{A}\right\rangle _{2}.\nonumber
\end{align}

But
\begin{align}
\vartheta^{\mathbf{b}}\wedge\left\langle A\vartheta^{\mathbf{a}}%
D_{e_{\mathbf{b}}}\widetilde{A}\right\rangle _{1}  &  =\left\langle
\vartheta^{\mathbf{b}}A\vartheta^{\mathbf{a}}D_{e_{\mathbf{b}}}\widetilde
{A}\right\rangle _{2}-\vartheta^{\mathbf{b}}\lrcorner\left\langle
A\vartheta^{\mathbf{a}}D_{e_{\mathbf{b}}}\widetilde{A}\right\rangle
_{3}=-\left\langle A\vartheta^{\mathbf{a}}D_{e_{\mathbf{b}}}\widetilde
{A}\right\rangle _{1}\wedge\vartheta^{\mathbf{b}}\nonumber\\
&  =-\left\langle A\vartheta^{\mathbf{a}}\left(  D_{e_{\mathbf{b}}}%
\widetilde{A}\right)  \vartheta^{\mathbf{b}}\right\rangle _{2}+\left\langle
A\vartheta^{\mathbf{a}}\left(  D_{e_{\mathbf{b}}}\widetilde{A}\right)
\right\rangle _{3}\llcorner\vartheta^{\mathbf{b}}\nonumber\\
&  =-\left\langle A\vartheta^{\mathbf{a}}\widetilde{%
\bpartial
A}\right\rangle _{2}+\left\langle A\vartheta^{\mathbf{a}}D_{e_{\mathbf{b}}%
}\widetilde{A}\right\rangle _{3}\llcorner\vartheta^{\mathbf{b}}\nonumber\\
&  =-\left\langle A\vartheta^{\mathbf{a}}\widetilde{F}\right\rangle
_{2}+\left\langle A\vartheta^{\mathbf{a}}D_{e_{\mathbf{b}}}\widetilde
{A}\right\rangle _{3}\llcorner\vartheta^{\mathbf{b}}=\left\langle
A\vartheta^{\mathbf{a}}F\right\rangle _{2}+\left\langle A\vartheta
^{\mathbf{a}}D_{e_{\mathbf{b}}}\widetilde{A}\right\rangle _{3}\llcorner
\vartheta^{\mathbf{b}} \label{J5}%
\end{align}
Then from Eq. (\ref{J5}) we can write Eq. (\ref{J4}) as
\begin{equation}
d\left(  A\vartheta^{\mathbf{a}}A\right)  =2\left\langle \left(
A\vartheta^{\mathbf{a}}\right)  F\right\rangle _{2}+2\vartheta^{\mathbf{b}%
}\lrcorner\left\langle A\vartheta^{\mathbf{a}}D_{e_{\mathbf{b}}}A\right\rangle
_{3} \label{J7}%
\end{equation}
and then
\begin{equation}
\mathcal{W}^{\mathbf{a}}=-\frac{1}{4}d(A\vartheta^{\mathbf{a}}A+X^{\mathbf{a}%
})=-\frac{1}{2}\left\langle A\vartheta^{\mathbf{a}}F\right\rangle _{2}%
-\frac{1}{2}B^{\mathbf{a}} \label{J8}%
\end{equation}

We now verify that $\underset{\mathtt{\eta}}{\delta}\mathcal{W}^{\mathbf{a}%
}=-\mathcal{T}^{\mathbf{a}}$. Indeed, since $\langle\vartheta^{\mathbf{b}%
}\langle A\vartheta^{\mathbf{a}}(D_{e_{\mathbf{b}}}F)\rangle_{4}\rangle
_{1}=0=\langle\vartheta^{\mathbf{b}}\langle(D_{e_{\mathbf{b}}}F)\vartheta
^{\mathbf{a}}A\rangle_{4}\rangle_{1}$ , taking into account the last identity
in Eq.(\ref{identities}) we can write
\begin{align}
\underset{\mathtt{\eta}}{\delta}\mathcal{W}^{\mathbf{a}}  &  =-%
\bpartial
\underset{\mathtt{\eta}}{{\lrcorner}}\mathcal{W}^{\mathbf{a}}=\frac{1}{2}%
\bpartial
\underset{\mathtt{\eta}}{{\lrcorner}}\left\langle \left(  A\vartheta
^{\mathbf{a}}\right)  F\right\rangle _{2}+\frac{1}{2}%
\bpartial
\lrcorner B^{\mathbf{a}}=\frac{1}{2}\left\langle
\bpartial
\left(  A\vartheta^{\mathbf{a}}F\right)  \right\rangle _{1}-\frac{1}{2}%
\bpartial
\left\langle \left(  A\vartheta^{\mathbf{a}}F\right)  \right\rangle _{0}%
+\frac{1}{2}%
\bpartial
\lrcorner B^{\mathbf{a}}\nonumber\\
&  =\frac{1}{2}\left\langle F\vartheta^{\mathbf{a}}F\right\rangle _{1}%
+\frac{1}{2}\left\langle \vartheta^{\mathbf{b}}A\vartheta^{\mathbf{a}%
}(D_{e_{\mathbf{b}}}F)\right\rangle _{1}\nonumber\\
&  =\frac{1}{2}\left\langle F\vartheta^{\mathbf{a}}F\right\rangle _{1}%
+\frac{1}{2}\left\langle \vartheta^{\mathbf{b}}\langle A\vartheta^{\mathbf{a}%
}(D_{e_{\mathbf{b}}}F)\rangle_{0}+\vartheta^{\mathbf{b}}\langle A\vartheta
^{\mathbf{a}}(D_{e_{\mathbf{b}}}F)\rangle_{2}+\vartheta^{\mathbf{b}}\langle
A\vartheta^{\mathbf{a}}(D_{e_{\mathbf{b}}}F)\rangle_{4}\right\rangle
_{1}\nonumber\\
&  =-\frac{1}{2}\left\langle F\vartheta^{\mathbf{a}}\widetilde{F}\right\rangle
_{1}+\frac{1}{2}\vartheta^{\mathbf{b}}\langle(D_{e_{\mathbf{b}}}%
F)\vartheta^{\mathbf{a}}A\rangle_{0}+\vartheta^{\mathbf{b}}\langle
(D_{e_{\mathbf{b}}}F)\vartheta^{\mathbf{a}}A\rangle_{2}-\frac{1}{2}%
\langle\vartheta^{\mathbf{b}}\langle(D_{e_{\mathbf{b}}}F)\vartheta
^{\mathbf{a}}A\rangle_{4}\rangle_{1}\nonumber\\
&  =-\frac{1}{2}F\vartheta^{\mathbf{a}}\widetilde{F}+\frac{1}{2}\left\langle
\vartheta^{\mathbf{b}}(D_{e_{\mathbf{b}}}F)\vartheta^{\mathbf{a}%
}A\right\rangle _{1}=-\frac{1}{2}F\vartheta^{\mathbf{a}}\widetilde{F}+\frac
{1}{2}\left\langle (%
\bpartial
F)\vartheta^{\mathbf{a}}A\right\rangle _{1}\nonumber\\
&  =-\frac{1}{2}F\vartheta^{\mathbf{a}}\widetilde{F}. \label{J9}%
\end{align}
Finally note that from Eqs. (\ref{J2}) and (\ref{J9}) we can write
($\mathcal{W}^{\mathbf{a}}=d\Gamma^{\mathbf{a}}=-\frac{1}{4}d\left(
A\vartheta^{\mathbf{a}}A+X^{\mathbf{a}}\right)  $ and $F=dA$)%
\begin{equation}%
\begin{array}
[c]{c}%
d\mathcal{W}^{\mathbf{a}}=0\text{, }\underset{\mathtt{\eta}}{\delta
}\mathcal{W}^{\mathbf{a}}=-\mathcal{T}^{\mathbf{a}},\\
\text{or}\\%
\bpartial
\mathcal{W}^{\mathbf{a}}=\mathcal{T}^{\mathbf{a}}.
\end{array}
\label{J10}%
\end{equation}
and we get the non trivial result that the $2$-form fields $\mathcal{W}%
^{\mathbf{a}}$ ($\mathbf{a=}0,1,2,3$)describing the energy-momentum
propagation satisfy Maxwell like equations $%
\bpartial
$ $\mathcal{W}^{\mathbf{a}}=\mathcal{T}^{\mathbf{a}}$ with sources being the
energy-momentum $1$-form fields $\mathcal{T}^{\mathbf{a}}$.

\section{The Effective Weitzenb\"{o}ck Spacetime Generated by $F$}

First we recall some results of \cite{edrod,quiroro,rodcap} where we showed
how effective Riemann-Cartan spacetimes can be generated by the presence of
distortion fields which arises from diffeomorphisms \texttt{h}: $M\rightarrow
M$. Here we investigate a particular diffeomorphism $\mathtt{h}$ associated
with the electromagnetic potential $A\in\sec\bigwedge\nolimits^{1}T^{\ast
}M\hookrightarrow\sec\mathcal{C\ell(}M,\mathtt{\eta}\mathcal{)}$ ($F=dA$) by%
\begin{equation}
\theta^{\mathbf{a}}=\mathtt{h}^{\ast}\Gamma^{\mathbf{a}}, \label{WE1}%
\end{equation}%
\begin{equation}
\mathbf{g}=\mathtt{h}^{\ast}\mathbf{\eta,} \label{g}%
\end{equation}%
\begin{equation}
\mathbf{g}=\eta_{\mathbf{ab}}\theta^{\mathbf{a}}\otimes\theta^{\mathbf{b}}%
\in\sec T_{2}^{0}M. \label{WE2}%
\end{equation}

Now, the metric of the cotangent bundle associated with $\mathbf{g}$ is
$g\in\sec T_{0}^{2}M$ and we have $g(\theta^{\mathbf{a}},\theta^{\mathbf{b}%
})=\eta^{\mathbf{ab}}$. Associated with $g$ there exists an extensor field
\cite{rodcap,femoro} $h:\sec\bigwedge\nolimits^{1}T^{\ast}M\longrightarrow
\sec\bigwedge\nolimits^{1}T^{\ast}M$ \ such that the extensor associated with
$g$ is \ \texttt{g }$=h^{2}:\sec\bigwedge\nolimits^{1}T^{\ast}M\longrightarrow
\sec\bigwedge\nolimits^{1}T^{\ast}M$ and we have
\begin{align}
\mathtt{g}(\theta^{\mathbf{a}},\theta^{\mathbf{b}})  &  =\theta^{\mathbf{a}%
}\underset{\mathtt{g}}{\cdot}\theta^{\mathbf{b}}=h(\theta^{\mathbf{a}%
})\underset{\mathtt{\eta}}{\cdot}h(\theta^{\mathbf{b}})\label{WE3}\\
&  =h(\mathtt{h}^{\ast}\Gamma^{\mathbf{a}})\underset{\mathtt{\eta}}{\cdot
}h(\mathtt{h}^{\ast}\Gamma^{\mathbf{a}})=\eta^{\mathbf{ab}}.
\end{align}

Modulus a Lorentz transformation we can take $h(\mathtt{h}^{\ast}%
\Gamma^{\mathbf{a}})=\vartheta^{\mathbf{a}}$, i.e.,%
\begin{equation}
h^{-1}(\vartheta^{\mathbf{a}})=\mathtt{h}^{\ast}\Gamma^{\mathbf{a}}%
=\theta^{\mathbf{a}}. \label{WE6}%
\end{equation}

Now we return to Eq.(\ref{J2}) , i.e., $\mathcal{W}^{\mathbf{a}}%
=d\Gamma^{\mathbf{a}}$ and write:%
\begin{equation}
\mathtt{h}^{\ast}\mathcal{W}^{\mathbf{a}}=\mathtt{h}^{\ast}d\Gamma
^{\mathbf{a}}=d\mathtt{h}^{\ast}\Gamma^{\mathbf{a}}. \label{WE7}%
\end{equation}

Calling
\[
\Theta^{\mathbf{a}}=\mathtt{h}^{\ast}\mathcal{W}^{\mathbf{a}}\in\sec
\bigwedge\nolimits^{2}T^{\ast}M
\]
we can write the last equation as
\begin{equation}
d\theta^{\mathbf{a}}=\Theta^{\mathbf{a}}. \label{WE8}%
\end{equation}

Now, recall that for any $\omega_{p}\in\sec\bigwedge\nolimits^{p}T^{\ast}M$ ,
if $\mathbf{g}=\mathtt{h}^{\ast}\mathbf{\eta}$ it holds that (see, e.g.,
\cite{rodcap})
\begin{equation}
\underset{\mathtt{g}}{\star}(\mathtt{h}^{\ast}\omega_{p})=\mathtt{h}^{\ast
}\underset{\mathtt{\eta}}{(\star}\omega_{p}), \label{WE9}%
\end{equation}
then, returning to Eq.(\ref{the equation}), equivalent to $d\underset
{\mathtt{\eta}}{\star}\mathcal{W}^{\mathbf{a}}=-\underset{\mathtt{\eta}}%
{\star}\mathcal{T}^{\mathbf{a}}$ we can write
\begin{equation}
d\underset{\mathtt{g}}{\star}(\mathtt{h}^{\ast}\mathcal{W}^{\mathbf{a}%
})=-\mathtt{h}^{\ast}(\underset{\mathtt{\eta}}{\star}\mathcal{T}^{\mathbf{a}%
})=-\underset{\mathtt{g}}{\star}(\mathtt{h}^{\ast}\mathcal{T}^{\mathbf{a}}).
\label{WE10}%
\end{equation}
Calling
\begin{equation}
T^{\mathbf{a}}=\mathtt{h}^{\ast}\mathcal{T}^{\mathbf{a}}\in\sec\bigwedge
\nolimits^{1}T^{\ast}M \label{WE T}%
\end{equation}
the deformed energy-momentum $1$-form field we have $d\underset{\mathtt{g}%
}{\star}\Theta^{\mathbf{a}}=-\underset{\mathtt{g}}{\star}T^{\mathbf{a}}$ or
equivalently%
\begin{equation}
\underset{\mathtt{g}}{\delta}\Theta^{\mathbf{a}}=-T^{\mathbf{a}}. \label{11}%
\end{equation}

We summarize the above results in the following :

\begin{proposition}
A free electromagnetic field $F\in\sec\bigwedge\nolimits^{2}T^{\ast}M$ living
in Minkowski spacetime structure $(M,\eta,D,\uparrow,\tau_{\mathtt{\eta}})$
and satisfying Maxwell equation $\mbox{\boldmath$\partial$}F=0$ generates an
effective Weitzenb\"{o}ck (or teleparallel) geometry , i..e., a teleparallel
spacetime $(M,g,\nabla,\uparrow,\tau_{\mathtt{g}})$ where $g=h^{\ast}\eta$,
$\nabla\mathbf{g}=0$, and the torsion $2$-forms $\Theta^{\mathbf{a}\text{ }}$
are given by Eq.(\ref{WE8}) in the teleparallel $g$-orthonormal cobasis
$\{\theta^{\mathbf{a}}\}$. Moreover the torsion $2$-form fields in this theory
propagate, i.e., satisfy Maxwell equations
\begin{equation}
d\Theta^{\mathbf{a}}=0\text{, }\underset{\mathtt{g}}{\delta}\Theta
^{\mathbf{a}}=-T^{\mathbf{a}}, \label{WE12}%
\end{equation}
with the surprising result that the sources of the $\Theta^{\mathbf{a}}$ are
the energy-momerntum $1$-form fields.
\end{proposition}

\section{The Effective Lorentzian Spacetime Generated by a Free
Electromagnetic Field $F$}

We now introduce an effective Lorentzian spacetime $(M,\mathbf{g}%
,\overset{\mathbf{g}}{D},\uparrow,\tau_{\mathtt{g}})$ as follows. We know that
for an electromagnetic field the trace of the energy-momentum tensor is null,
i.e., $\mathcal{T}_{\mathbf{a}}^{\mathbf{a}}=0$. It follows that with
$\mathbf{g=}\mathtt{h}^{\ast}\eta$ as defined above we have that the trace of
$T^{\mathbf{a}}=$ $\mathtt{h}^{\ast}\mathcal{T}^{\mathbf{a}}$ is also null.
Then since the Einstein equations can be written \ (see, e.g.,
\cite{rodcap,edrod}) as
\begin{equation}
\underset{\mathtt{g}}{\star}\mathcal{R}^{\mathbf{a}}-\frac{1}{2}%
R\underset{\mathtt{g}}{\star}\theta^{\mathbf{a}}=-\underset{\mathtt{g}}{\star
}T^{\mathbf{a}}, \label{L1}%
\end{equation}
where $\mathcal{R}^{\mathbf{a}}=R_{\mathbf{b}}^{\mathbf{a}}\theta^{\mathbf{b}%
}\in\sec\bigwedge\nolimits^{1}T^{\ast}M$ $\hookrightarrow\mathcal{C\ell
}(M,\mathtt{g})$ are the \textit{Ricci} $1$-form fields and $R=R_{\mathbf{a}%
}^{\mathbf{a}}$ is the scalar curvature, which for the present case is null
since $R=-T_{\mathbf{a}}^{\mathbf{a}}=0$.

Then, we have that
\begin{equation}
\underset{\mathtt{g}}{\star}\mathcal{R}^{\mathbf{a}}=-\underset{\mathtt{g}%
}{\star}T^{\mathbf{a}}. \label{L2}%
\end{equation}

Using the fact that $d\underset{\mathtt{g}}{\star}\Theta^{\mathbf{a}%
}=-\underset{\mathtt{g}}{\star}T^{\mathbf{a}}$ (where now, of course, we are
not interpreting the $\Theta^{\mathbf{a}}$ as $2$-forms of torsion of a
teleparallel connection) we get
\begin{equation}
\underset{\mathtt{g}}{\star}\mathcal{R}^{\mathbf{a}}=d\underset{\mathtt{g}%
}{\star}\Theta^{\mathbf{a}}, \label{L3}%
\end{equation}
or equivalently
\begin{equation}
\underset{\mathtt{g}}{\delta}\Theta^{\mathbf{a}}=\mathcal{R}^{\mathbf{a}}
\label{L4}%
\end{equation}

Moreover recalling that the \textit{Hodge Laplacian} is $\Diamond
=-(d\underset{\mathtt{g}}{\delta}+\underset{\mathtt{g}}{\delta}d)$ we
have\footnote{We mention also that the Ricci $1$-forms may be written in terms
of the Ricci operator $()$}
\begin{equation}
\mathcal{R}^{\mathbf{a}}=-\Diamond\theta^{\mathbf{a}}-d\underset{\mathtt{g}%
}{\delta}\theta^{\mathbf{a}} \label{L5}%
\end{equation}
or taking into account Eqs. (\ref{WE1}), (\ref{L2}) (\ref{WE T}) we end with
\[
\Diamond\theta^{\mathbf{a}}+d\underset{\mathtt{g}}{\delta}\theta^{\mathbf{a}%
}=\frac{1}{2}\underset{\mathtt{g}}{\star}\left(  \mathtt{h}^{\ast}%
(F\vartheta^{\mathbf{a}}F)\right)  .
\]

We summarize the above results in the following

\begin{proposition}
A free electromagnetic field $F\in\sec\bigwedge\nolimits^{2}T^{\ast}M$ living
in Minkowski spacetime structure $(M,\eta,D,\uparrow,\tau_{\mathtt{\eta}})$
and satisfying Maxwell equation $%
\bpartial
F=0$ generates an effective Lorentzian $(M,\mathbf{g},\overset{\mathbf{g}}%
{D},\uparrow,\tau_{\mathtt{g}})$ where $\mathbf{g}=h^{\ast}\mathbf{\eta}$,
$\overset{\mathbf{g}}{D}\mathbf{g}=0$, such that the dual of the Ricci
$1$-forms are exact differentials, i.e., $\underset{\mathtt{g}}{\star
}\mathcal{R}^{\mathbf{a}}=-d\underset{\mathtt{g}}{\star}\Theta^{\mathbf{a}}$.
Moreover, we have
\begin{equation}
\mathcal{R}^{\mathbf{a}}=-\Diamond\theta^{\mathbf{a}}-d\underset{\mathtt{g}%
}{\delta}\theta^{\mathbf{a}}. \label{L6}%
\end{equation}

\end{proposition}

\begin{remark}
\emph{Eq.(\ref{L6}) }is the condition that the Ricci $1$-form fields in a
Lorentzian spacetime modelling a gravitational field must satisfy in order to
describe an electromagnetic field propagating in a Minkowski spacetime. We
then arrive in a kind of already unified theory as in the
Rainich-Misner-Wheeler theory.
\end{remark}

\begin{remark}
Recalling that we can write
\begin{equation}
-\underset{\mathtt{g}}{\star}\mathcal{R}^{\mathbf{a}}=d\underset{\mathtt{g}%
}{\star}\mathcal{S}^{\mathbf{a}}+\underset{\mathtt{g}}{\star}t^{\mathbf{a}%
}=\underset{\mathtt{g}}{\star}T^{\mathbf{a}}, \label{l7}%
\end{equation}
where\footnote{See, e.g., \cite{thiwal,rodcap}.}
\begin{align}
\underset{\mathtt{g}}{\star}t_{\mathbf{\ }}^{\mathbf{c}}  &  =-\frac{1}{2}%
\mbox{\boldmath{$\omega$}}%
_{\mathbf{ab}}\wedge\lbrack%
\mbox{\boldmath{$\omega$}}%
_{\mathbf{d}}^{\mathbf{c}}\wedge\underset{\mathtt{g}}{\star}(\theta
^{\mathbf{a}}\wedge\theta^{\mathbf{b}}\wedge\theta^{\mathbf{d}})+%
\mbox{\boldmath{$\omega$}}%
_{\mathbf{d}}^{\mathbf{b}}\wedge\underset{\mathtt{g}}{\star}(\theta
^{\mathbf{a}}\wedge\theta^{\mathbf{d}}\wedge\theta^{\mathbf{c}})],\nonumber\\
\underset{\mathtt{g}}{\star}\mathcal{S}^{\mathbf{c}}  &  =\frac{1}{2}%
\mbox{\boldmath{$\omega$}}%
_{\mathbf{ab}}\wedge\underset{\mathtt{g}}{\star}(\theta^{\mathbf{a}}%
\wedge\theta^{\mathbf{b}}\wedge\theta^{\mathbf{c}}). \label{7.10.17}%
\end{align}
with $\omega^{\mathbf{cd}}\in\sec%
{\displaystyle\bigwedge\nolimits^{1}}
T^{\ast}M\hookrightarrow\sec\mathcal{C\ell}(M,\mathtt{g})\mathcal{\ }$given
~by
\begin{equation}
\omega^{\mathbf{cd}}=\frac{1}{2}\left[  \theta^{\mathbf{d}}\underset
{\mathtt{g}}{\lrcorner}d\theta^{\mathbf{c}}-\theta^{\mathbf{c}}\underset
{\mathtt{g}}{\lrcorner}d\theta^{\mathbf{d}}+\theta^{\mathbf{c}}\underset
{\mathtt{g}}{\lrcorner}\left(  \theta^{\mathbf{d}}\underset{\mathtt{g}%
}{\lrcorner}d\theta_{\mathbf{a}}\right)  \theta^{\mathbf{a}}\right]  .
\end{equation}
which recalling that $\theta^{\mathbf{a}}=\mathtt{h}^{\ast}\Gamma^{\mathbf{a}%
}$ and that \cite{rodcap} $\mathtt{h}^{\ast}\omega_{p}\underset{\mathtt{g}%
}{\lrcorner}\mathtt{h}^{\ast}\omega_{k}=$ $\mathtt{h}^{\ast}(\omega
_{p}\underset{\mathtt{\eta}}{\lrcorner}\omega_{k})$ for any $\omega_{p}\in\sec%
{\displaystyle\bigwedge\nolimits^{p}}
T^{\ast}M$, $\omega_{k}\in\sec%
{\displaystyle\bigwedge\nolimits^{k}}
T^{\ast}M$ can also be written as%

\begin{equation}
\omega^{\mathbf{cd}}=\frac{1}{2}\mathtt{h}^{\ast}\left[  \Gamma^{\mathbf{d}%
}\underset{\mathtt{\Gamma}}{\lrcorner}d\Gamma^{\mathbf{c}}-\Gamma^{\mathbf{c}%
}\underset{\mathtt{\eta}}{\lrcorner}d\Gamma^{\mathbf{d}}+\Gamma^{\mathbf{c}%
}\underset{\mathtt{\eta}}{\lrcorner}\left(  \Gamma^{\mathbf{d}}\underset
{\mathtt{\eta}}{\lrcorner}d\Gamma_{\mathbf{a}}\right)  \Gamma^{\mathbf{a}%
}\right]
\end{equation}
Using \emph{Eq.(\ref{WE1})} we can rewrite \emph{Eqs.(\ref{7.10.17})} as
\begin{align}
\underset{\mathtt{g}}{\star}t_{\mathbf{\ }}^{\mathbf{c}}  &  =-\frac{1}%
{2}\mbox{\boldmath{$\omega$}}_{\mathbf{ab}}\wedge\lbrack
\mbox{\boldmath{$\omega$}}_{\mathbf{d}}^{\mathbf{c}}\wedge\underset
{\mathtt{g}}{\star}(\theta^{\mathbf{a}}\wedge\theta^{\mathbf{b}}\wedge
\theta^{\mathbf{d}})+\mbox{\boldmath{$\omega$}}_{\mathbf{d}}^{\mathbf{b}%
}\wedge\underset{\mathtt{g}}{\star}(\theta^{\mathbf{a}}\wedge\theta
^{\mathbf{d}}\wedge\theta^{\mathbf{c}})]\nonumber\\
&  =\frac{1}{4}\mathtt{h}^{\ast}\left[  \Gamma_{\mathbf{b}}\underset
{\mathtt{\eta}}{\lrcorner}d\Gamma_{\mathbf{a}}-\Gamma_{\mathbf{a}}%
\underset{\mathtt{\eta}}{\lrcorner}d\Gamma_{\mathbf{b}}+\Gamma_{\mathbf{a}%
}\underset{\mathtt{\eta}}{\lrcorner}\left(  \Gamma_{\mathbf{b}}\underset
{\mathtt{\eta}}{\lrcorner}d\Gamma_{\mathbf{p}}\right)  \Gamma^{\mathbf{p}%
}\right] \nonumber\\
&  \;\;\qquad\wedge\mathtt{h}^{\ast}\left[  \Gamma_{\mathbf{d}}\underset
{\mathtt{\eta}}{\lrcorner}d\Gamma^{\mathbf{c}}-\Gamma^{\mathbf{c}}%
\underset{\mathtt{\eta}}{\lrcorner}d\Gamma_{\mathbf{d}}+\Gamma^{\mathbf{c}%
}\underset{\mathtt{\eta}}{\lrcorner}\left(  \Gamma^{\;\mathbf{d}}%
\underset{\mathtt{\eta}}{\lrcorner}d\Gamma_{\mathbf{p}}\right)  \Gamma
^{\mathbf{p}}\right] \nonumber\\
&  {}\;\;\qquad\wedge\mathtt{h}^{\ast}\underset{\mathtt{\eta}}{\star}%
(\Gamma^{\mathbf{a}}\wedge\Gamma^{\mathbf{b}}\wedge\Gamma^{\mathbf{d}%
})\nonumber\\
&  \quad+\frac{1}{2}\wedge\mathtt{h}^{\ast}\left[  \Gamma_{\mathbf{d}%
}\underset{\mathtt{\eta}}{\lrcorner}d\Gamma^{\mathbf{b}}-\Gamma^{\mathbf{b}%
}\underset{\mathtt{\eta}}{\lrcorner}d\Gamma_{\mathbf{d}}+\Gamma^{\mathbf{b}%
}\underset{\mathtt{\eta}}{\lrcorner}\left(  \Gamma^{\;\mathbf{d}}%
\underset{\mathtt{\eta}}{\lrcorner}d\Gamma_{\mathbf{p}}\right)  \Gamma
^{\mathbf{p}}\right]  \wedge\mathtt{h}^{\ast}\underset{\mathtt{\eta}}{\star
}(\Gamma^{\mathbf{a}}\wedge\Gamma^{\mathbf{b}}\wedge\Gamma^{\mathbf{d}%
}),\nonumber\\
\underset{\mathtt{g}}{\star}\mathcal{S}^{\mathbf{c}}  &  =\frac{1}%
{4}\mathtt{h}^{\ast}\left[  \Gamma_{\mathbf{b}}\underset{\mathtt{\eta}%
}{\lrcorner}d\Gamma_{\mathbf{a}}-\Gamma_{\mathbf{a}}\underset{\mathtt{\eta}%
}{\lrcorner}d\Gamma_{\mathbf{b}}+\Gamma_{\mathbf{a}}\underset{\mathtt{\eta}%
}{\lrcorner}\left(  \Gamma_{\mathbf{b}}\underset{\mathtt{\eta}}{\lrcorner
}d\Gamma_{\mathbf{p}}\right)  \Gamma^{\mathbf{p}}\right]  \wedge
\mathtt{h}^{\ast}\underset{\mathtt{\eta}}{\star}(\Gamma^{\mathbf{a}}%
\wedge\Gamma^{\mathbf{b}}\wedge\Gamma^{\mathbf{d}}). \label{eqr}%
\end{align}

\end{remark}

Finally taking into account that $\underset{\mathtt{g}}{\star}t_{\mathbf{\ }%
}^{\mathbf{c}}=\underset{\mathtt{g}}{\star}\mathtt{h}^{\ast}t_{M\mathbf{\ }%
}^{\mathbf{c}}=\mathtt{h}^{\ast}\underset{\mathtt{\eta}}{\star}t_{M\mathbf{\ }%
}^{\mathbf{c}}$ and \ $\underset{\mathtt{g}}{\star}\mathcal{S}_{\mathbf{\ }%
}^{\mathbf{c}}=\underset{\mathtt{g}}{\star}\mathtt{h}^{\ast}S_{M\mathbf{\ }%
}^{\mathbf{c}}=\mathtt{h}^{\ast}\underset{\mathtt{\eta}}{\star}S_{M\mathbf{\ }%
}^{\mathbf{c}}$ we get%

\begin{align}
\underset{\mathtt{\eta}}{\star}t_{M\mathbf{\ }}^{\mathbf{c}}  &  =\frac{1}%
{4}\left[  \Gamma_{\mathbf{b}}\underset{\mathtt{\eta}}{\lrcorner}%
d\Gamma_{\mathbf{a}}-\Gamma_{\mathbf{a}}\underset{\mathtt{\eta}}{\lrcorner
}d\Gamma_{\mathbf{b}}+\Gamma_{\mathbf{a}}\underset{\mathtt{\eta}}{\lrcorner
}\left(  \Gamma_{\mathbf{b}}\underset{\mathtt{\eta}}{\lrcorner}d\Gamma
_{\mathbf{p}}\right)  \Gamma^{\mathbf{p}}\right] \nonumber\\
&  \;\;\qquad\wedge\left[  \Gamma_{\mathbf{d}}\underset{\mathtt{\eta}%
}{\lrcorner}d\Gamma^{\mathbf{c}}-\Gamma^{\mathbf{c}}\underset{\mathtt{\eta}%
}{\lrcorner}d\Gamma_{\mathbf{d}}+\Gamma^{\mathbf{c}}\underset{\mathtt{\eta}%
}{\lrcorner}\left(  \Gamma^{\;\mathbf{d}}\underset{\mathtt{\eta}}{\lrcorner
}d\Gamma_{\mathbf{p}}\right)  \Gamma^{\mathbf{p}}\right] \nonumber\\
&  {}\;\;\qquad\wedge\underset{\mathtt{\eta}}{\star}(\Gamma^{\mathbf{a}}%
\wedge\Gamma^{\mathbf{b}}\wedge\Gamma^{\mathbf{d}})\nonumber\\
&  \quad+\frac{1}{2}\wedge\left[  \Gamma_{\mathbf{d}}\underset{\mathtt{\eta}%
}{\lrcorner}d\Gamma^{\mathbf{b}}-\Gamma^{\mathbf{b}}\underset{\mathtt{\eta}%
}{\lrcorner}d\Gamma_{\mathbf{d}}+\Gamma^{\mathbf{b}}\underset{\mathtt{\eta}%
}{\lrcorner}\left(  \Gamma^{\;\mathbf{d}}\underset{\mathtt{\eta}}{\lrcorner
}d\Gamma_{\mathbf{p}}\right)  \Gamma^{\mathbf{p}}\right]  \wedge
\underset{\mathtt{\eta}}{\star}(\Gamma^{\mathbf{a}}\wedge\Gamma^{\mathbf{b}%
}\wedge\Gamma^{\mathbf{d}}),\nonumber\\
\underset{\mathtt{\eta}}{\star}\mathcal{S}_{M}^{\mathbf{c}}  &  =\frac{1}%
{4}\left[  \Gamma_{\mathbf{b}}\underset{\mathtt{\eta}}{\lrcorner}%
d\Gamma_{\mathbf{a}}-\Gamma_{\mathbf{a}}\underset{\mathtt{\eta}}{\lrcorner
}d\Gamma_{\mathbf{b}}+\Gamma_{\mathbf{a}}\underset{\mathtt{\eta}}{\lrcorner
}\left(  \Gamma_{\mathbf{b}}\underset{\mathtt{\eta}}{\lrcorner}d\Gamma
_{\mathbf{p}}\right)  \Gamma^{\mathbf{p}}\right]  \wedge\underset
{\mathtt{\eta}}{\star}(\Gamma^{\mathbf{a}}\wedge\Gamma^{\mathbf{b}}%
\wedge\Gamma^{\mathbf{d}}).
\end{align}
To end we recall that since $d\underset{\mathtt{g}}{\star}T^{\mathbf{a}}=0$ we
also have$\ d\underset{\mathtt{g}}{\star}t^{\mathbf{a}}=0$. Then, in our
theory the $\Theta^{\mathbf{a}}$ are also superpotentials for the
gravitational field described by the tetrad fields $\{\theta^{\mathbf{a}}\}$!

\section{Conclusions}

In this paper we proved using the Clifford bundle formalism that the
energy-momentum densities $\underset{\mathtt{\eta}}{\star}\mathcal{T}%
^{\mathbf{a}}=\frac{1}{2}\underset{\mathtt{\eta}}{\star}(F\vartheta
^{\mathbf{a}}\tilde{F})$ of a free electromagnetic field $F$ living on
Minkowski spacetime are sources of exact \ $2$-forms $\mathcal{W}^{\mathbf{a}%
}$ which satisfy Maxwell like equations $d\mathcal{W}^{\mathbf{a}}=0$,
$d\underset{\mathtt{\eta}}{\star}\mathcal{W}^{\mathbf{a}}=\underset
{\mathtt{\eta}}{\star}\mathcal{T}^{\mathbf{a}}$. Which this noticeable result
\ we show that the free electromagnetic field may be interpreted as generating
a (teleparallel) Weitzenb\"{o}ck spacetime [$(M,\mathbf{g},\nabla
,\uparrow,\tau_{\mathtt{g}})$] \ or an effective Lorentzian spacetime
[$(M,\mathbf{g},\overset{\mathbf{g}}{D},\uparrow,\tau_{\mathtt{g}})$], whose
properties are determined with details. In both structures \ the metric
$\mathbf{g=}\mathtt{h}^{\ast}\mathbf{\eta}$ \ where $\mathtt{h}:M\rightarrow
M$ is a conformal diffeomorphism given by Eq.(\ref{g}).

\section*{Acknowledgement}

E. A. N. C. thanks the Direcci\'{o}n de Investigaci\'{o}n de la Universidad de
La Serena DIULS, for parcial financial support. Rold\~{a}o da Rocha thanks the
Funda\c{c}\~{a}o de Amparo \`{a} Pesquisa do Estado de S\~{a}o Paulo - Brazil
(FAPESP) for financial support (PDJ - 05/03071-0).

\appendix

\section{Clifford Bundles}

Let $(M,\mathbf{\eta},D,\tau_{\mathtt{\eta}},\uparrow)$ be Minkowski
spacetime. $(M,\mathbf{\eta})$ is a four dimensional space oriented (by the
volume form $\tau_{\mathtt{\eta}})$ and time oriented (by the equivalence
relation $\uparrow$, see \cite{rodcap}) Lorentzian manifold, with
$M\simeq\mathbb{R}^{4}$ and $\mathbf{\eta}\in\sec T_{2}^{0}M$ is a Lorentzian
metric of signature $(1,3)$. $T^{\ast}M$ [$TM$] is the cotangent [tangent]
bundle. $T^{\ast}M=\cup_{x\in M}T_{x}^{\ast}M$, $TM=\cup_{x\in M}T_{x}M$, and
$T_{x}M\simeq T_{x}^{\ast}M\simeq\mathbb{R}^{1,3}$, where $\mathbb{R}^{1,3}$
is the Minkowski vector space~. $D$ is the Levi-Civita connection of
$\mathbf{\eta}$, $i.e$\textit{.\/}, $D\mathbf{\eta}=0$, $\mathbf{R}(D)=0$.
Also $\Theta(D)=0$, $\mathbf{R}$ and $\Theta$ being respectively the torsion
and curvature tensors. Let $\mathtt{\eta}\in\sec T_{0}^{2}M$ be the metric on
the cotangent bundle associated with $\mathbf{\eta}\in\sec T_{2}^{0}M$. The
Clifford bundle of differential forms $\mathcal{C}\!\ell(M,\mathtt{\eta})$ is
the bundle of algebras, i.e., $\mathcal{C}\!\ell(M,\mathtt{\eta})=\cup_{x\in
M}\mathcal{C}\!\ell(T_{x}^{\ast}M)$, where $\forall x\in M$, $\mathcal{C}%
\!\ell(T_{x}^{\ast}M)=\mathbb{R}_{1,3}$, the so called \emph{spacetime}
\emph{algebra}. Recall also that $\mathcal{C}\!\ell(M,\mathtt{\eta})$ is a
vector bundle associated with the \emph{\ }$\mathbf{\eta}$-\emph{orthonormal
frame bundle \ }$\mathbf{P}_{\mathrm{SO}_{(1,3)}^{e}}(M,\mathbf{\eta})$, i.e.,
$\mathcal{C}\!\ell(M,\mathtt{\eta})$ $=P_{SO_{+(1,3)}}(M)\times_{ad}%
\mathbb{R}_{1,3}$ (see more details in, e.g., \cite{mosnawal,rodcap}). For any
$x\in M$, $\mathcal{C}\!\ell(T_{x}^{\ast}M)$ is a linear space over the real
field $\mathbb{R}$. Moreover, $\mathcal{C}\!\ell(T_{x}^{\ast}M)$ is isomorphic
to the Cartan algebra $\bigwedge T_{x}^{\ast}M$ of the cotangent space and
$\bigwedge T_{x}^{\ast}M=\sum_{k=0}^{4}\bigwedge{}^{k}T_{x}^{\ast}M$, where
$\bigwedge^{k}T_{x}^{\ast}M$ is the $\binom{4}{k}$-dimensional space of
$k$-forms. Then, sections of $\mathcal{C}\!\ell(M,\mathtt{\eta})$ can be
represented as a sum of non homogeneous differential forms. Let $\{x^{\mu}\}$
be coordinates in Einstein-Lorentz-Poincar\'{e} gauge for $M$ and let
$\{e_{\mu}=\partial/\partial x^{\mu}\}\in\sec FM$ (the frame bundle) be an
orthonormal basis for $TM$, i.e., $\mathbf{\eta}(e_{\mu},e_{\nu})=\eta_{\mu
\nu}=\mathrm{diag}(1,-1,-1,-1)$, Let $\gamma^{\nu}=dx^{\nu}\in\sec
\bigwedge^{1}T^{\ast}M\hookrightarrow\sec\mathcal{C}\!\ell(M,\mathbf{\eta})$
($\nu=0,1,2,3$) such that the set $\{\gamma^{\nu}\}$ is the dual basis of
$\{e_{\mu}\},$ and of course, \ $\mathtt{\eta}(\gamma^{\mu},\gamma^{\nu}%
)=\eta^{\mu\nu}=\mathrm{diag}(1,-1,-1,-1)$. \ We introduce moreover the
notations $\vartheta^{\mathbf{a}}=\delta_{\mu}^{\mathbf{a}}dx^{\mu}$ and
$e_{\mathbf{a}}=\delta_{\mathbf{a}}^{\mu}\frac{\partial}{\partial x^{\mu}}$.
We say that $\{e_{\mathbf{a}}\}$ is a section of the orthonormal frame bundle
$\mathbf{P}_{\mathrm{SO}_{(1,3)}^{e}}(M,\mathbf{\eta})$ and its dual basis
$\{\vartheta^{\mathbf{a}}\}$ is a section of the orthonormal coframe bundle
(denoted $P_{\mathrm{SO}_{(1,3)}^{e}}(M,\mathbf{\eta})$).

\subsection{Clifford Product}

The fundamental \emph{Clifford product} (in what follows to be denoted by
juxtaposition of symbols) is generated by $\theta^{\mathbf{a}}\theta
^{\mathbf{b}}+\theta^{\mathbf{b}}\theta^{\mathbf{a}}=2\eta^{\mathbf{ab}}$ and
if $\mathcal{C}\in\mathcal{C}\!\ell(M,\mathtt{\eta})$ we have%

\begin{equation}
\mathcal{C}=s+v_{\mathbf{a}}\theta^{\mathbf{a}}+\frac{1}{2!}b_{\mathbf{ab}%
}\theta^{\mathbf{a}}\theta^{\mathbf{b}}+\frac{1}{3!}a_{\mathbf{abc}}%
\theta^{\mathbf{a}}\theta^{\mathbf{b}}\theta^{\mathbf{c}}+p\theta^{5}\;,
\label{3}%
\end{equation}
where $\tau_{\mathtt{\eta}}:=\theta^{5}=\theta^{0}\theta^{1}\theta^{2}%
\theta^{3}=dx^{0}dx^{1}dx^{2}dx^{3}$ is the volume element and $s$,
$v_{\mathbf{a}}$, $b_{\mathbf{ab}}$, $a_{\mathbf{abc}}$, $p\in\sec
\bigwedge^{0}T^{\ast}M\hookrightarrow\sec\mathcal{C}\!\ell(M,\mathtt{\eta})$.

Let $\mathcal{A}_{r},\in\sec\bigwedge^{r}T^{\ast}M\hookrightarrow
\sec\mathcal{C}\!\ell(M,\mathtt{\eta}),\mathcal{B}_{s}\in\sec\bigwedge
^{s}T^{\ast}M\hookrightarrow\sec\mathcal{C}\!\ell(M,\mathtt{\eta})$. For
$r=s=1$, we define the \emph{scalar product} as follows:

For $a,b\in\sec\bigwedge^{1}T^{\ast}M\hookrightarrow\sec\mathcal{C}%
\!\ell(M,\mathtt{\eta}),$%
\begin{equation}
a\underset{\mathtt{\eta}}{\cdot}b=\frac{1}{2}(ab+ba)=\mathtt{\eta}(a,b).
\label{4}%
\end{equation}
We define also the \emph{exterior product} ($\forall r,s=0,1,2,3)$ by
\begin{align}
\mathcal{A}_{r}\wedge\mathcal{B}_{s}  &  =\langle\mathcal{A}_{r}%
\mathcal{B}_{s}\rangle_{r+s},\nonumber\\
\mathcal{A}_{r}\wedge\mathcal{B}_{s}  &  =(-1)^{rs}\mathcal{B}_{s}%
\wedge\mathcal{A}_{r}, \label{5}%
\end{align}
where $\langle\rangle_{k}$ is the \textit{component} in $\bigwedge^{k}T^{\ast
}M$ \ (projection) of the Clifford field. The exterior product is extended by
linearity to all sections of $\mathcal{C}\!\ell(M,\mathtt{\eta})$.

For $\mathcal{A}_{r}=a_{1}\wedge...\wedge a_{r},$ $\mathcal{B}_{r}=b_{1}%
\wedge...\wedge b_{r}$, the scalar product is defined here as follows,
\begin{align}
\mathcal{A}_{r}\underset{\mathtt{\eta}}{\cdot}\mathcal{B}_{r}  &
=(a_{1}\wedge...\wedge a_{r})\underset{\mathtt{\eta}}{\cdot}(b_{1}%
\wedge...\wedge b_{r})\nonumber\\
&  =\left\vert
\begin{array}
[c]{lll}%
a_{1}\underset{\mathtt{\eta}}{\cdot}b_{1} & .... & a_{1}\underset
{\mathtt{\eta}}{\cdot}b_{r}\\
.......... & .... & ..........\\
a_{r}\underset{\mathtt{\eta}}{\cdot}b_{1} & .... & a_{r}\underset
{\mathtt{\eta}}{\cdot}b_{r}%
\end{array}
\right\vert . \label{6}%
\end{align}

We agree that if $r=s=0$, the scalar product is simple the ordinary product in
the real field.

Also, if $r\neq s$, then $\mathcal{A}_{r}\cdot\mathcal{B}_{s}=0$. Finally, the
scalar product is extended by linearity for all sections of $\mathcal{C}%
\!\ell(M,\mathtt{\eta})$.

For $r\leq s,$ $\mathcal{A}_{r}=a_{1}\wedge...\wedge a_{r},$ $\mathcal{B}%
_{s}=b_{1}\wedge...\wedge b_{s\text{ }}$we define the left contraction by
\begin{equation}
\underset{\mathtt{\eta}}{\llcorner}:(\mathcal{A}_{r},\mathcal{B}_{s}%
)\mapsto\mathcal{A}_{r}\underset{\mathtt{\eta}}{\lrcorner}\mathcal{B}_{s}=%
{\displaystyle\sum\limits_{i_{1}\,<...\,<i_{r}}}
\epsilon^{i_{1}....i_{s}}(a_{1}\wedge...\wedge a_{r})\underset{\mathtt{\eta}%
}{\cdot}(b_{_{i_{1}}}\wedge...\wedge b_{i_{r}})^{\sim}b_{i_{r}+1}%
\wedge...\wedge b_{i_{s}} \label{7}%
\end{equation}
where $\sim$ is the reverse mapping (\emph{reversion}) defined by
\begin{equation}
\sim:\sec\bigwedge^{p}T^{\ast}M\ni a_{1}\wedge...\wedge a_{p}\mapsto
a_{p}\wedge...\wedge a_{1} \label{8}%
\end{equation}
and extended by linearity to all sections of $\mathcal{C}\!\ell(M,\mathtt{\eta
})$. We agree that for $\alpha,\beta\in\sec\bigwedge^{0}T^{\ast}M$ the
contraction is the ordinary (pointwise) product in the real field and that if
$\alpha\in\sec\bigwedge^{0}T^{\ast}M\hookrightarrow\mathcal{C}\!\ell
(M,\mathtt{\eta})$, $\mathcal{A}_{r}\in\sec\bigwedge^{r}T^{\ast}%
M\hookrightarrow\mathcal{C}\!\ell(M,\mathtt{\eta})$, $\mathcal{B}_{s}\in
\sec\bigwedge^{s}T^{\ast}M$ $\hookrightarrow\mathcal{C}\!\ell(M,\mathtt{\eta
})$\ then $(\alpha\mathcal{A}_{r})\underset{\mathtt{\eta}}{\lrcorner
}\mathcal{B}_{s}=\mathcal{A}_{r}\underset{\mathtt{\eta}}{\lrcorner}%
(\alpha\mathcal{B}_{s})$. Left contraction is extended by linearity to all
pairs of elements of sections of $\mathcal{C}\!\ell(M,\mathtt{\eta})$, i.e.,
for $\mathcal{A},\mathcal{B}\in\sec\mathcal{C}\!\ell(M,\mathtt{\eta})$%

\begin{equation}
\mathcal{A\underset{\mathtt{\eta}}{\lrcorner}B}=\sum_{r,s}\langle
\mathcal{A}\rangle_{r}\underset{\mathtt{\eta}}{\lrcorner}\langle
\mathcal{B}\rangle_{s},\text{ }r\leq s. \label{9}%
\end{equation}

It is also necessary to introduce the operator of \emph{right contraction}
denoted by $\llcorner$. The definition is obtained from the one presenting the
left contraction with the imposition that $r\geq s$ and taking into account
that now if $\mathcal{A}_{r}\in\sec\bigwedge^{r}T^{\ast}M\hookrightarrow
\mathcal{C}\!\ell(M,\mathtt{\eta})$, $\mathcal{B}_{s}\in\sec\bigwedge
^{s}T^{\ast}M\hookrightarrow\mathcal{C}\!\ell(M,\mathtt{\eta})\ $then
$\mathcal{A}_{r}\underset{\mathtt{\eta}}{\llcorner}(\alpha\mathcal{B}%
_{s})=(\alpha\mathcal{A}_{r})\underset{\mathtt{\eta}}{\llcorner}%
\mathcal{B}_{s}$.

The main formulas used in the present paper can be obtained (details may be
found in \cite{rodcap}) from the following ones (where $a\in\sec\bigwedge
^{1}T^{\ast}M\hookrightarrow\sec\mathcal{C}\!\ell(M,\mathtt{\eta})$):
\begin{align}
a\mathcal{B}_{s}  &  =a\underset{\mathtt{\eta}}{\lrcorner}\mathcal{B}%
_{s}+a\wedge\mathcal{B}_{s},\mathcal{B}_{s}a=\mathcal{B}_{s}\underset
{\mathtt{\eta}}{\llcorner}a+\mathcal{B}_{s}\wedge a,\nonumber\\
a\underset{\mathtt{\eta}}{\lrcorner}\mathcal{B}_{s}  &  =\frac{1}%
{2}(a\mathcal{B}_{s}-(-1)^{s}\mathcal{B}_{s}a),\nonumber\\
\mathcal{A}_{r}\underset{\mathtt{\eta}}{\lrcorner}\mathcal{B}_{s}  &
=(-1)^{r(s-r)}\mathcal{B}_{s}\llcorner\mathcal{A}_{r},\nonumber\\
a\wedge\mathcal{B}_{s}  &  =\frac{1}{2}(a\mathcal{B}_{s}+(-1)^{s}%
\mathcal{B}_{s}a),\nonumber\\
\mathcal{A}_{r}\mathcal{B}_{s}  &  =\langle\mathcal{A}_{r}\mathcal{B}%
_{s}\rangle_{|r-s|}+\langle\mathcal{A}_{r}\mathcal{B}_{s}\rangle
_{|r-s|+2}+...+\langle\mathcal{A}_{r}\mathcal{B}_{s}\rangle_{|r+s|}\nonumber\\
&  =\sum\limits_{k=0}^{m}\langle\mathcal{A}_{r}\mathcal{B}_{s}\rangle
_{|r-s|+2k}\text{ }\nonumber\\
\mathcal{A}_{r}\underset{\mathtt{\eta}}{\cdot}\mathcal{B}_{r}  &
=\mathcal{B}_{r}\underset{\mathtt{\eta}}{\cdot}\mathcal{A}_{r}=\widetilde
{\mathcal{A}}_{r}\text{ }\underset{\mathtt{\eta}}{\lrcorner}\mathcal{B}%
_{r}=\mathcal{A}_{r}\underset{\mathtt{\eta}}{\llcorner}\widetilde{\mathcal{B}%
}_{r}=\langle\widetilde{\mathcal{A}}_{r}\mathcal{B}_{r}\rangle_{0}%
=\langle\mathcal{A}_{r}\widetilde{\mathcal{B}}_{r}\rangle_{0},
\end{align}

\begin{align}
\langle\mathcal{AB}\rangle_{r}  &  =(-1)^{r(r-1)/2}\langle\widetilde
{\mathcal{B}}\widetilde{\mathcal{A}}\rangle_{r},\nonumber\\
\langle\mathcal{A}_{r}\mathcal{B}_{s}\rangle_{r}  &  =\langle\widetilde
{\mathcal{B}}_{s}A_{r}\rangle_{r}=(-1)^{s(s-1)/2}\langle\mathcal{B}%
_{s}\mathcal{A}_{r}\rangle_{r},\nonumber\\
\langle\mathcal{A}_{r}\mathcal{B}_{s}\mathcal{C}_{t}\rangle_{q}  &
=(-1)^{\varepsilon}\langle\mathcal{C}_{t}\mathcal{B}_{s}\mathcal{A}_{r}%
\rangle_{q},\nonumber\\
\varepsilon &  =\frac{1}{2}(q^{2}+r^{2}+s^{2}+t^{2}-q-r-s-t)
\label{identities}%
\end{align}

\subsection{Hodge Star Operator $\underset{\mathtt{\eta}}{\star}$}

Let $\underset{\mathtt{\eta}}{\star}$ be the Hodge star operator, i.e., the
mapping
\[
\underset{\mathtt{\eta}}{\star}:\bigwedge^{k}T^{\ast}M\rightarrow
\bigwedge^{4-k}T^{\ast}M,\text{ }\mathcal{A}_{k}\mapsto\underset{\mathtt{\eta
}}{\star}\mathcal{A}_{k}%
\]
where for $\mathcal{A}_{k}\in\sec\bigwedge^{k}T^{\ast}M\hookrightarrow
\mathcal{C}\!\ell(M,\mathtt{\eta})$%
\begin{equation}
\lbrack\mathcal{B}_{k}\underset{\mathtt{\eta}}{\cdot}\mathcal{A}_{k}%
]\tau_{\mathtt{\eta}}=\mathcal{B}_{k}\wedge\underset{\mathtt{\eta}}{\star
}\mathcal{A}_{k},\text{ }\forall\mathcal{B}_{k}\in\sec\bigwedge\nolimits^{k}%
T^{\ast}M\hookrightarrow\sec\mathcal{C}\!\ell(M,\mathtt{\eta}). \label{11a}%
\end{equation}
$\tau_{\mathtt{\eta}}=\theta^{5}\in\bigwedge^{4}T^{\ast}M$ is a
\emph{standard} volume element. Then we can verify that
\begin{equation}
\underset{\mathtt{\eta}}{\star}\mathcal{A}_{k}=\widetilde{\mathcal{A}}_{k}%
\tau_{\mathtt{\eta}}=\widetilde{\mathcal{A}}_{k}\theta^{5}. \label{11b}%
\end{equation}

\subsection{$\mathcal{C\ell}(M,\mathtt{\eta})$, $\mathcal{C\ell}%
(M,\mathtt{g})$, \texttt{h} and $h$}

\subsubsection{Enter \texttt{h}}

In this section \texttt{h} is a diffeomorphis $\mathtt{h}:M\rightarrow M$,
$\mathfrak{e}\mapsto\mathtt{h}\mathfrak{e}$ such that if $\mathtt{h}^{\ast
}\Gamma^{\mathbf{a}}=\theta^{\mathbf{a}}$ and if $\mathbf{\eta\in}\sec
T_{2}^{0}M$ then%

\begin{equation}
\mathtt{\mathbf{g}}=\mathtt{h}^{\ast}\mathbf{\eta=\eta_{\mathbf{ab}}}%
\theta^{\mathbf{a}}\otimes\theta^{\mathbf{b}}\mathbf{,} \label{10hypa}%
\end{equation}
\medskip where $\{\theta^{\mathbf{a}}\}$ is the $\mathbf{g}$-orthonormal
cobasis used in the main text.

\subsubsection{Enter $h$}

Consider the Clifford bundles of nonhomogeneous multiform fields
$\mathcal{C\ell}(M,\mathtt{\eta})$ and $\mathcal{C\ell}(M,\mathtt{g})$. In
$\mathcal{C\ell}(M,\mathtt{\eta})$ we denoted the Clifford product by
juxtaposition of symbols, the scalar product by $\underset{\mathtt{\eta}%
}{\cdot}$ and the contractions by $\underset{\mathtt{\eta}}{\lrcorner}$ and
$\underset{\mathtt{\eta}}{\llcorner}$ and by $\underset{\mathtt{\eta}}{\star}$
we denote the Hodge dual. The Clifford product in $\mathcal{C\ell
}(M,\mathtt{g})$ will be denoted by the symbol $\vee$, the scalar product will
be denoted by $\bullet\equiv\underset{\mathtt{g}}{\cdot}$ and the contractions
by $\underset{\mathtt{\mathbf{g}}}{\lrcorner}$ and $\underset
{\mathtt{\mathbf{g}}}{\llcorner}$ while by $\underset{\mathtt{\mathbf{g}}%
}{\star}$ we denote the Hodge dual operator associated with
\texttt{$\mathbf{g}$}.

Let $\{\mathbf{e}_{\mathbf{a}}\}$ be a non coordinate basis of $TM$ dual to
the cobasis $\{\theta^{\mathbf{a}}\}$. We take the $\theta^{\mathbf{a}}$ as
sections of the Clifford bundle $\mathcal{C}\ell(M,\mathtt{\eta})$, i.e.,
$\theta^{\mathbf{a}}\in\sec\bigwedge\nolimits^{1}T^{\ast}M\hookrightarrow
\sec\mathcal{C}\ell(M,\mathtt{\eta})$. In this basis taking into account
Eq.(\ref{10hypa}) that $\mathtt{g}$\texttt{ }$\in\sec T_{0}^{2}M$ is given by
\begin{equation}
\mathtt{g}=\eta^{\mathbf{ab}}\mathbf{e}_{\mathbf{a}}\otimes\mathbf{e}%
_{\mathbf{b}}. \label{8.1a}%
\end{equation}
The cobasis $\{\vartheta^{\mathbf{a}}\}$ defines a Clifford product in
$\mathcal{C}\ell(M,\mathtt{\eta})$ by
\begin{equation}
\vartheta^{\mathbf{a}}\vartheta^{\mathbf{b}}+\vartheta^{\mathbf{b}}%
\vartheta^{\mathbf{a}}=2\eta^{\mathbf{ab}}, \label{8.3}%
\end{equation}
and taking into account that the cobasis $\{\theta^{\mathbf{a}}\}$ defines a
\textit{deformed} Clifford product $\vee$ in $\mathcal{C}\ell(M,\mathtt{\eta
})$ (see details in \cite{rodcap}) generating a representation of the Clifford
bundle $\mathcal{C\ell}(M,\mathtt{g})$ we can write
\begin{align}
\theta^{\mathbf{a}}\vee\theta^{\mathbf{b}}  &  =\theta^{\mathbf{a}}%
\bullet\theta^{\mathbf{b}}+\theta^{\mathbf{a}}\wedge\theta^{\mathbf{b}%
},\nonumber\\
\theta^{\mathbf{a}}\vee\theta^{\mathbf{b}}+\theta^{\mathbf{b}}\vee
\theta^{\mathbf{a}}  &  =2\eta^{\mathbf{ab}}. \label{8.4}%
\end{align}
Then, as proved, e.g., in \cite{rodcap} there exist $(1,1)$-extensor fields
$g$ and $h^{-1}$ such that
\begin{equation}
\mathtt{g}(\theta^{\mathbf{a}},\theta^{\mathbf{b}})=\theta^{\mathbf{a}}%
\bullet\theta^{\mathbf{b}}=\theta^{\mathbf{a}}\cdot g(\theta^{\mathbf{b}%
})=h(\theta^{\mathbf{a}})\cdot h(\theta^{\mathbf{b}})=\eta^{\mathbf{ab}}.
\label{8.5}%
\end{equation}

The gauge metric extensor $h:\sec%
{\displaystyle\bigwedge\nolimits^{1}}
T^{\ast}M\rightarrow\sec%
{\displaystyle\bigwedge\nolimits^{1}}
T^{\ast}M$ is defined (modulus a Lorentz transformation) by
\begin{equation}
h(\theta^{\mathbf{a}})=\vartheta^{\mathbf{a}}. \label{8.5new}%
\end{equation}

\subsubsection{Relation Between $h$ and $\mathtt{h}^{\ast}$}

Introduce coordinates functions $\{%
\sly
^{\mu}\}$ in the Einstein-Lorentz-Poincar\'{e} gauge for $M$ such that $%
\sly
^{\mu}=$ $%
\slx
^{\mu}$ (where the $\{%
\slx
^{\mu}\}$ are the coordinate functions in the Einstein -Lorentz-Poincar\'{e}
gauge already introduced above). Put
\begin{align}%
\slx
^{\mu}(\mathfrak{e})  &  =x^{\mu},%
\slx
^{\mu}(\mathtt{h}\mathfrak{e})=y^{\mu},\nonumber\\%
\sly
^{\mu}(\mathtt{h}\mathfrak{e)}  &  =%
\slx
^{\mu}(\mathtt{h}\mathfrak{e})=y^{\mu}. \label{8.27'}%
\end{align}
If $y^{\mu}=\mathtt{h}^{\mu}(x^{\nu})$ is the coordinate
expression\footnote{Recall that the $\mathtt{h}^{\mu}$ are invertible
differentiable functions.} for \texttt{h} , the coordinate expressions of
$\mathbf{\eta}$ at \texttt{h}$\mathfrak{e}$ and $\mathtt{\mathbf{g}}$ at
$\mathfrak{e}$ can be written as:%
\begin{equation}
\left.  \mathbf{\eta}\right\vert _{\mathtt{h}\mathfrak{e}}\mathbf{=}%
\eta_{\mathbf{ab}}\delta_{\mu}^{\mathbf{a}}\delta_{\nu}^{\mathbf{b}%
}dy^{\mathbf{\mu}}\otimes dy^{\nu}\text{, }\left.  \mathtt{\mathbf{g}%
}\right\vert _{\mathfrak{e}}=g_{\mu\nu}dx^{\mu}\otimes dx^{\nu}. \label{8.27a}%
\end{equation}
Since $\left.  \mathtt{h}^{\ast}\mathbf{\eta(}\delta_{\mathbf{a}}^{\mu
}\mathbf{\partial/\partial}x^{\mu},\delta_{\mathbf{b}}^{\nu}\mathbf{\partial
/\partial}x^{\nu}\mathbf{)}\right\vert _{\mathfrak{e}}\mathbf{=}\left.
\mathbf{\eta(}\delta_{\mathbf{a}}^{\mu}\mathtt{h}_{\ast}\mathbf{\partial
/\partial}x^{\mu},\delta_{\mathbf{b}}^{\nu}\mathtt{h}_{\ast}\mathbf{\partial
/\partial}x^{\nu}\mathbf{)}\right\vert _{\mathtt{h}\mathfrak{e}}$ we have
\begin{equation}
\mathtt{\mathbf{g}}=\mathtt{h}^{\ast}\mathbf{\eta}=\eta_{\mathbf{ab}}%
\delta_{\alpha}^{\mathbf{a}}\delta_{\beta}^{\mathbf{b}}\frac{\partial
y^{\alpha}}{\partial x^{\mu}}\frac{\partial y^{\beta}}{\partial x^{\nu}%
}dx^{\mu}\otimes dx^{\nu} \label{8.27b}%
\end{equation}
with
\begin{equation}
g_{\mu\nu}=\eta_{\mathbf{ab}}\delta_{\alpha}^{\mathbf{a}}\delta_{\beta
}^{\mathbf{b}}\frac{\partial y^{\alpha}}{\partial x^{\mu}}\frac{\partial
y^{\beta}}{\partial x^{\nu}}. \label{8.27c}%
\end{equation}

Now, take notice that at $\mathfrak{e}$,$\ \{\mathbf{f}_{\mathbf{a}}\}$,
$\mathbf{f}_{\mathbf{a}}=\delta_{\mathbf{a}}^{\mu}\mathtt{h}_{\ast}%
^{-1}\mathbf{\partial/\partial}y^{\mu}=\delta_{\mathbf{a}}^{\mu}\frac{\partial
x^{\nu}}{\partial y^{\mu}}\frac{\partial}{\partial x^{\nu}}$ is not (in
general) a coordinate basis for $TM$. It is also not $\mathbf{\eta}%
$-orthonormal\footnote{Indeed, $\mathbf{\eta}(\mathbf{e}_{\mathbf{a}%
},\mathbf{e}_{\mathbf{b}})=\delta_{\mathbf{a}}^{\mu}\delta_{\mathbf{b}}^{\nu
}\frac{\partial y^{\alpha}}{\partial x^{\mu}}\frac{\partial y^{\beta}%
}{\partial x^{\nu}}\eta_{\alpha\beta}$.}. The dual basis of $\{\mathbf{f}%
_{\mathbf{a}}\}$ at $\mathfrak{e}$ is $\{\left.  \sigma^{\mathbf{a}%
}\right\vert _{\mathfrak{e}}\}$\textbf{, }with\textbf{ }$\left.
\sigma^{\mathbf{a}}\right\vert _{\mathfrak{e}}=\left.  \delta_{\mu
}^{\mathbf{a}}\frac{\partial y^{\mu}}{\partial x^{\nu}}dx^{\nu}\right\vert
_{\mathfrak{e}}=\mathtt{h}^{\ast}(\left.  \delta_{\mu}^{\mathbf{a}}dy^{\mu
}\right\vert _{\mathtt{h}\mathfrak{e}})$. Then it exists an extensor field
$\check{h}\ $differing from $h$ by a Lorentz extensor, i.e., $\check
{h}=h\Lambda$ such that$\ \left.  \sigma^{\mathbf{a}}\right\vert
_{\mathfrak{e}}=\left.  \check{h}^{-1}(\delta_{\mu}^{\mathbf{a}}dy^{\mu
})\right\vert _{\mathfrak{e}}=\left.  \check{h}_{\mu}^{-1\mathbf{a}}dy^{\mu
}\right\vert _{\mathfrak{e}}$, where, we have for any $\mathfrak{e}\in M$,
$\ $%
\begin{equation}
\delta_{\alpha}^{\mathbf{a}}\frac{\partial y^{\alpha}}{\partial x^{\mu}%
}=\check{h}_{\mu}^{-1\mathbf{a}}. \label{8.27dd}%
\end{equation}
$\ \ $To determine $\check{h}$ we proceed as follows. Suppose
$\mathtt{\mathbf{g}}=\eta_{\mathbf{ab}}\sigma^{\mathbf{a}}\otimes
\sigma^{\mathbf{b}}$ is known. Let ($v_{i},\lambda_{i})$ be respectively the
eigen-covectors and the eigenvalues of $g$, i.e., $g(v_{i})=\lambda_{i}v_{i}$
(no sum in $i$) and $\{\vartheta^{\mathbf{a}}\}$ the $\eta$-orthonormal
coordinate basis for $T^{\ast}M$ introduced above. Then, since $g=\check
{h}^{2}=h^{2}$ we immediately have
\begin{equation}
\check{h}(v_{i})=\sqrt{\left\vert \lambda_{i}\right\vert }\mathtt{\eta}%
(v_{i},\vartheta_{\mathbf{a}})\vartheta^{\mathbf{a}}, \label{8.29bis}%
\end{equation}
which then determines the extensor field $h$ (modulus a \ local Lorentz
rotation) at any spacetime point, and thus the diffeomorphism \texttt{h}
(modulus a local Lorentz rotation).

\subsubsection{Relation Between $\underset{\mathtt{\eta}}{\star}$ and
$\underset{\mathtt{g}}{\star}$}

If $g=h^{2}$ we have that \cite{femoro,rodcap} for any $\omega_{p}\in
\sec\bigwedge T^{\ast}M$
\begin{equation}
\underset{\mathtt{g}}{\star}\omega_{p}=\ \underline{h}\text{ }\underset
{\mathtt{\eta}}{\star}\underline{h}\omega_{p}, \label{hodges}%
\end{equation}
where $\ \underline{h}$ is the exterior power extension of $h$.

\subsection{Dirac Operator acting on Sections of a General Clifford Bundle
$\mathcal{C}\ell(M,\mathtt{g})$}

Let $d$ and $\underset{\mathtt{g}}{\delta}$ be respectively the differential
and Hodge codifferential operators acting on sections of $\sec\bigwedge
\nolimits^{k}T^{\ast}M\hookrightarrow\sec\mathcal{C}\ell(M,\mathtt{\eta})$. If
$\mathcal{A}_{p}\in\sec\bigwedge^{p}T^{\ast}M\hookrightarrow\sec
\mathcal{C}\ell(M,\mathtt{g})$, then $\underset{\mathtt{g}}{\delta}%
\mathcal{A}_{p}=(-1)^{p}\underset{\mathtt{g}}{\star}^{-1}d\underset
{\mathtt{g}}{\star}\mathcal{A}_{p}$, with $\underset{\mathtt{g}}{\star}%
^{-1}\underset{\mathtt{g}}{\star}=\mathrm{identity}$.

The Dirac operator acting on sections of $\mathcal{C}\!\ell(M,\mathtt{g})$ is
the invariant first order differential operator
\begin{equation}
\mbox{\boldmath$\partial$}=\theta^{\mathbf{a}}\overset{\mathbf{g}}%
{D}_{\mathbf{e}_{\mathbf{a}}}, \label{12}%
\end{equation}
where $\overset{\mathbf{g}}{D}_{\mathbf{e}_{\mathbf{a}}}$ is the Levi-Civita
connection of $\mathbf{g}$.

\subsubsection{Useful Formula for Calculation of $\overset{\mathbf{g}}%
{D}_{\mathbf{e}_{\mathbf{a}}}\mathcal{A}$}

The \textit{reciprocal }basis of $\{\theta^{\mathbf{b}}\}$ is denoted
$\{\theta_{\mathbf{a}}\}$ and we have $\theta_{\mathbf{a}}\underset
{\mathtt{g}}{\cdot}\theta_{\mathbf{b}}=\eta_{\mathbf{ab}}$ ($\eta
_{\mathbf{ab}}=\mathrm{diag}(1,-1,-1,-1)$). Also,
\begin{equation}
\overset{\mathbf{g}}{D}_{\mathbf{e}_{\mathbf{a}}}\theta^{\mathbf{b}}%
=-\omega_{\mathbf{ac}}^{\mathbf{b}}\theta^{\mathbf{c}}=-\omega_{\mathbf{a}%
}^{\mathbf{bc}}\theta_{\mathbf{c}}, \label{12n}%
\end{equation}
with $\omega_{\mathbf{a}}^{\mathbf{bc}}=-\omega_{\mathbf{a}}^{\mathbf{cb}}$,
and $\omega_{\mathbf{a}}^{\mathbf{bc}}=\eta^{\mathbf{bk}}\omega_{\mathbf{kal}%
}\eta^{\mathbf{cl}},$ $\omega_{\mathbf{abc}}=\eta_{\mathbf{ad}}\omega
_{\mathbf{bc}}^{\mathbf{d}}=-\omega_{\mathbf{cba}}$. Defining
\begin{equation}
\mathbf{\omega}_{\mathbf{a}}=\frac{1}{2}\omega_{\mathbf{a}}^{\mathbf{bc}%
}\theta_{\mathbf{b}}\wedge\theta_{\mathbf{c}}\in\sec%
{\displaystyle\bigwedge\nolimits^{2}}
T^{\ast}M\hookrightarrow\sec\mathcal{C}\!\ell(M,\mathtt{g}), \label{12nn}%
\end{equation}
we have (by linearity) that \cite{mosnawal} for any $\mathcal{A}\in
\sec\bigwedge T^{\ast}M\hookrightarrow\sec\mathcal{C}\!\ell(M,\mathtt{g})$
\begin{equation}
\overset{\mathbf{g}}{D}_{\mathbf{e}_{\mathbf{a}}}\mathcal{A}=\partial
_{\mathbf{e}_{\mathbf{a}}}\mathcal{A}+\frac{1}{2}[\mathbf{\omega}_{\mathbf{a}%
},\mathcal{A}], \label{12nnn}%
\end{equation}
where $\partial_{\mathbf{e}_{\mathbf{a}}}$ is the Pfaff
derivative\footnote{E.g., if $A=\frac{1}{p!}A_{\mathbf{i}_{1}...\mathbf{i}%
_{p}}\theta^{_{\mathbf{i}_{1}}}...\theta^{_{.\mathbf{i}_{p}\text{ }}}$then
$\partial_{\mathbf{e}_{\mathbf{a}}}A=\frac{1}{p!}[\mathbf{e}_{\mathbf{a}%
}(A_{\mathbf{i}_{1}...\mathbf{i}_{p}})]\theta^{_{\mathbf{i}_{1}}}%
...\theta^{_{.\mathbf{i}_{p}\text{ }}}$.}

\subsection{Dirac Operator $\mbox{\boldmath$\partial$}=$ $d-$ $\underset
{\mathtt{g}}{\delta}$}

Using Eq.(\ref{12nnn}) we can easily show the very important result:%

\begin{align}
\mbox{\boldmath$\partial$}\vee\mathcal{A}  &
=\mbox{\boldmath$\partial$}\wedge\mathcal{A}%
+\,\mbox{\boldmath$\partial$}\underset{\mathtt{g}}{\lrcorner}\mathcal{A}%
=d\mathcal{A}-\underset{\mathtt{g}}{\delta}\mathcal{A},\nonumber\\
\mbox{\boldmath$\partial$}\wedge\mathcal{A}  &  =d\mathcal{A},\hspace
{0.1in}\,\mbox{\boldmath$\partial$}\underset{\mathtt{g}}{\lrcorner}%
\mathcal{A}=-\underset{\mathtt{g}}{\delta}\mathcal{A}. \label{13}%
\end{align}

\begin{remark}
We will use the symbol $%
\bpartial
$ for the Dirac operator acting on sections of $\mathcal{C}\!\ell
(M,\mathtt{\eta})$ over Minkowski spacetime. In this case we have with
$\{e_{\mathbf{a}}\}$ a $\mathbf{\eta}$-orthonormal basis and $\{\vartheta
^{\mathbf{a}}\}$ its dual basis (as defined above)%
\[%
\bpartial
=\vartheta^{\mathbf{a}}D_{e_{\mathbf{a}}}%
\]%
\begin{align}%
\bpartial
\mathcal{A}  &  =%
\bpartial
\wedge\mathcal{A}+\,%
\bpartial
\underset{\mathtt{\eta}}{\lrcorner}\mathcal{A}=d\mathcal{A}-\underset
{\mathtt{\eta}}{\delta}\mathcal{A},\nonumber\\%
\bpartial
\wedge\mathcal{A}  &  =d\mathcal{A},\hspace{0.1in}\,%
\bpartial
\underset{\mathtt{\eta}}{\lrcorner}\mathcal{A}=-\underset{\mathtt{\eta}%
}{\delta}\mathcal{A}. \label{14bis}%
\end{align}
\newline\linebreak
\end{remark}

\end{document}